\documentclass[pre,floatfix,twocolumn,showpacs]{revtex4}

\usepackage{amsmath}
\usepackage{amssymb}
\usepackage{graphicx}

\newcommand{\ie}{\emph{i.}$\,$\emph{e.}}
\newcommand{\eg}{\emph{e.}$\,$\emph{g.}}
\newcommand{\etal}{\emph{et}$\,$\emph{al.}}

\newcommand{\romd}{{\operatorname{d}}}

\newcommand{\romm}{{\operatorname{m}}}
\newcommand{\romp}{{\operatorname{p}}}
\newcommand{\romA}{{\text{A}}}
\newcommand{\romB}{{\text{B}}}

\newcommand{\VECa}{{\boldsymbol{a}}}
\newcommand{\VECe}{{\boldsymbol{e}}}
\newcommand{\VECf}{{\boldsymbol{f}}}
\newcommand{\VECl}{{\boldsymbol{l}}}
\newcommand{\VECm}{{\boldsymbol{m}}}
\newcommand{\VECn}{{\boldsymbol{n}}}
\newcommand{\VECr}{{\boldsymbol{r}}}
\newcommand{\VECt}{{\boldsymbol{t}}}
\newcommand{\VECx}{{\boldsymbol{x}}}
\newcommand{\VECy}{{\boldsymbol{y}}}
\newcommand{\VECz}{{\boldsymbol{z}}}

\newcommand{\VECF}{{\boldsymbol{F}}}
\newcommand{\VECX}{{\boldsymbol{X}}}

\newcommand{\VECnab}{{\boldsymbol{\nabla}}}

\newcommand{\RR}{\mathbb{R}}

\begin{document}

\title{Interface mediated interactions between particles -- a geometrical approach}

\author{Martin Michael M\"uller}
\author{Markus Deserno}
\affiliation{Max-Planck-Institut f\"ur Polymerforschung, %
             Ackermannweg 10, %
             55128 Mainz, %
             Germany}
\author{Jemal Guven}
\affiliation{Instituto de Ciencias Nucleares, %
             Universidad Nacional Aut\'onoma de M\'exico, %
             Apdo.\ Postal 70-543, %
             04510 M\'exico D.F., %
             Mexico}

\date{\today}
\begin{abstract}
  Particles bound to an interface interact because they deform its
  shape.  The stresses that result are fully encoded in the geometry
  and described by a divergence-free surface stress tensor.  This
  stress tensor can be used to express the force on a particle as a
  line integral along any conveniently chosen closed contour that
  surrounds the particle.  The resulting expression is exact (\ie,
  free of any ``smallness'' assumptions) and independent of the chosen
  surface parametrization. Additional surface degrees of freedom, such
  as vector fields describing lipid tilt, are readily included in this
  formalism.  As an illustration, we derive the exact force for
  several important surface Hamiltonians in various symmetric
  two-particle configurations in terms of the midplane geometry; its
  sign is evident in certain interesting limits.  Specializing to the
  linear regime, where the shape can be analytically determined, these
  general expressions yield force-distance relations, several of which
  have originally been derived by using an energy based approach.
\end{abstract}


\maketitle


\section{Introduction\label{sec:introduction}}

The interaction between spatially separated objects is mediated by the
disturbance of the region that surrounds them, described by a
\emph{field}.  In electromagnetic theory for example the interaction
between charged particles is described by the Maxwell field equations.
Since they are linear, interactions add.  However, more often than
not, the field equations are nonlinear as for example in the case of
General Relativity: even though the energy-momentum tensor couples
linearly to the curvature, the latter depends in a nonlinear way on
the spacetime metric and its derivatives \cite{MTW,Wald,Weinberg}.
The source of the nonlinearity lies in the geometric nature of the
problem.  Not only do interactions fail to add up, even the humble two
particle problem poses challenges.

``Effective'' interactions between macroscopic degrees of freedom
arise in statistical physics when a partial trace is performed in the
partition function over unobserved microscopic degrees of freedom
\cite{Belloni00,Likos01}.  The Boltzmann factor invariably renders these
interactions nonlinear. This time, the source of the nonlinearity is
the entropy hidden in the degrees of freedom that have been traced
out.  For example, the effective interaction between charged colloids
in salty water is described (at a mean-field level) by the
Poisson-Boltzmann equation \cite{Hill}.

In this paper we will discuss a classical example which belongs to the
class of effective interactions, while owing its nonlinearity to its
geometric origin: the interaction between particles mediated by the
deformation of a surface to which they are bound. This problem
includes the capillary interactions between particles bound to
liquid-fluid interfaces \cite{pinningexp,pinningtheo,Kralchevsky}, or
the membrane mediated interactions between colloids or proteins
adhering to or embedded in lipid bilayer membranes
\cite{Kralchevsky,Koltover,Weiklcyl,gbp,inclusions,Fourspher,Kim,BisBis}.
To approach the problem we require two pieces of information. First,
how does the energy of the surface depend on its shape, or in other
words, what is the ``surface Hamiltonian''? Second, how does a bound
particle locally deform the surface?  With this information, we may
(in principle) deduce the equilibrium shape minimizing the energy of
the surface for any given placement of the bound particles. Knowing
the shape, the energy can be determined by integration, and the forces
it implies follow by differentiating with respect to appropriate
placement variables. In general, however, the ground state of the
surface is a solution of a nonlinear field equation (``the shape
equation''), thereby thwarting progress by this route at a very early stage.

Sometimes the linearization of a nonlinear theory is adequate.  Just
as one recovers Newtonian gravitation as the weak-field limit of
General Relativity \cite{MTW,Wald,Weinberg}, or Debye-H\"uckel theory
as the weak-field limit of Poisson-Boltzmann theory \cite{Hill}, a
linear theory for surface mediated interactions is useful for certain
simple geometries, notably weakly perturbed flat surfaces.  At this
level, the approach to interactions based on energy becomes tractable.
Yet, linearization is also often inadequate.  The full theory may
display qualitatively new effects which are absent in the linearized
theory: strong gravitational fields give us black holes
\cite{MTW,Wald,Weinberg}; the bare charge of a highly charged colloid
gets strongly renormalized by counterion condensation
\cite{colloidchargeren}.

There is, however, an alternative approach to interactions which was
outlined in \cite{mem_inter}.  By relating the \emph{interaction}
between particles to the equilibrium \emph{geometry} of the surface, a
host of exact nonlinear results is provided. The link is formed by the
\emph{surface stress tensor}, and it can be established without
solving the shape equation. Once we know the Hamiltonian, we can
express the stress at any point in terms of the local geometry --
covariantly and without any approximation.  We will briefly revisit
the essentials of this construction in
Sec.~\ref{sec:surfacestresstensor}.  Knowing the stresses, the force
on a particle is then determined by a line integral of the stress
tensor along any surface contour enclosing the particle, as we will
show in Sec.~\ref{sec:forcesbetweenparticles}.

Such results might, at first sight, appear somewhat formal: without
the equilibrium surface shape, they cannot be translated into hard
numbers.  However, the close link between the force and the geometry,
combined with a very general knowledge one has about the surface shape
(\eg, its symmetry) will turn out to provide valuable qualitative
insight into the nature of the interaction (\eg, its sign).  Even on a
completely practical level, this approach scores points against the
traditional approach involving energy, providing a significantly more
efficient way to extract forces from the surface shape determined
numerically (in whatever way).

We will illustrate this approach with a selection of examples
involving different symmetries and surface Hamiltonians.  In
Sec.~\ref{sec:tilt} we demonstrate how its scope extends in a very
natural way to include internal degrees of freedom on the membrane --
in particular: a vector order parameter which has for instance been
used to describe lipid tilt
\cite{HelPro88,MacLub,NePo9293,SeShNe96,HamKoz00,tilt_force}.  To make
contact with the energy based approach in the literature, and also in
order to link the formalism to a more familiar setting, we specialize
in Sec.~\ref{sec:Monge} to a Monge parametrization and its
linearization.  This will permit us in Sec.~\ref{sec:examples} to
derive force-distance curves for interactions mediated by surface
tension, membrane curvature, and lipid tilt.  Various well-known
linear results \cite{pinningexp,pinningtheo,Weiklcyl} then follow very
naturally using the stress tensor approach.


\section{Energy from Geometry \label{sec:surfacestresstensor}}

In this paper we want to study the physics of interfaces, which are
characterized by a reparametrization invariant surface Hamiltonian.
The appropriate language for this is differential geometry, and in
this section we will outline how physical questions can be
formulated very efficiently in this language. We
first summarize the necessary mathematical basics and introduce our
notation (the reader will find more background material in
Refs.~\cite{DifferentialGeometry}). We then define the class of
Hamiltonians we will be considering. The corresponding
Euler-Lagrange equations for the interface degrees of freedom
will be cast as a conservation law. The most direct way to do this is
to implement all geometrical constraints using Lagrange
mulipliers; not only does this approach provide a quick derivation
of the shape equation, it also provides a transparent physical
identification of the surface stresses.

\subsection{Differential Geometry and Notation\label{subsec:diffgeoandnotation}}

We consider a two-dimensional surface $\Sigma$ embedded in
three-dimensional Euclidean space $\RR^3$, which is described locally
by its position $\VECX(\xi^1,\xi^2)\in\RR^3$, where the $\xi^a$ are a
suitable set of local coordinates on the surface.  The embedding
functions $\VECX$ induce two geometrical tensors which completely
describe the surface: the \emph{metric} $g_{ab}$ and the
\emph{extrinsic curvature} $K_{ab}$, defined by
\begin{subequations}
\label{eq:structuralrelationsship1}
\begin{eqnarray}
  g_{ab} & = & \VECe_{a} \cdot \VECe_{b} \qquad \text{and} \label{eq:gab} \\
  K_{ab} & = & \VECe_{a} \cdot \partial_{b}\VECn \ ,
  \label{eq:Kab}
\end{eqnarray}
\end{subequations}
where $a,b \in \{1,2\}$.  The local coordinate frame formed by the
tangent vectors $\VECe_{1}$ and $\VECe_{2}$ extended by the normal
vector $\VECn$ forms a local basis of $\RR^{3}$:
\begin{subequations}
\label{eq:structuralrelationsship2}
\begin{eqnarray}
  \VECe_{a} & = & \partial \VECX /\partial \xi^{a} \; = \; \partial_{a}\VECX \
  , \label{eq:ea} \\
  \VECe_{a} \cdot \VECn & = & 0  \ , \label{eq:ea_normal_n} \\
  \VECn^{2} & = & 1 \ . \label{eq:n2=1}
\end{eqnarray}
\end{subequations}
Note that unlike $\VECn$, the $\VECe_{a}$ are generally not
normalized.

In the following, $\nabla_a$ denotes the metric-compatible covariant
derivative \cite{covariant_derivative} and $\Delta=\nabla_a\nabla^a$
the corresponding Laplacian. Surface indices are raised with the
inverse metric $g^{ab}$.  The trace of the extrinsic curvature,
$K=g^{ab}K_{ab}$, is twice the mean curvature.  Using the above sign
conventions, a sphere of radius $a$ with \emph{outward} pointing unit
normal has a \emph{positive} $K=2/a$.

The intrinsic and extrinsic geometries are related by the
Gauss-Codazzi-Mainardi equations
\begin{subequations}
\label{eq:GCM}
\begin{eqnarray}
  \nabla_aK_{bc}\,-\,\nabla_bK_{ac} & = & 0 \ ,
  \label{eq:GCM1}
  \\
  K_{ac}K_{bd} - K_{ad}K_{bc} & = & R_{abcd} \ ,
  \label{eq:GCM2}
\end{eqnarray}
\end{subequations}
where $R_{abcd}$ is the Riemann tensor constructed with the metric;
its contraction over the first and third index is the Ricci tensor,
$R_{bd} = g^{ac}R_{abcd}$, whose further contraction gives the Ricci
scalar curvature, $R = g^{bd}R_{bd}$.  From Eqn.~(\ref{eq:GCM2}) we
see that the latter satisfies $R = K^2 - K^{ab}K_{ab}$.  In particular
in two dimensions we have that $R=2\,K_{\text{G}}$, where
$K_{\text{G}} = \det(K_a^b)$ is the Gaussian curvature (Gauss'
Theorema Egregium \cite{DifferentialGeometry}).

\subsection{Surface energy and its variation \label{subsec:surfaceenergetics}}

We consider surfaces such as lipid membranes and soap films,
characterized by the property that the associated energy is completely
determined by the surface geometry and described by a Hamiltonian
which is an integral of a local Hamiltonian density $\mathcal{H}$ over
the surface:
\begin{equation}
  H[\VECX] = \int_\Sigma \romd A \;
  \mathcal{H}(g_{ab},K_{ab},\nabla_a K_{bc}, \ldots) \ .
  \label{eq:surfacefunctional}
\end{equation}
The infinitesimal area element is $\romd A = \sqrt{g} \, \romd^2
\xi$, where $g=\det(g_{ab})$ is the determinant of the metric. The
density $\mathcal{H}$ depends only on scalars constructed from local
surface tensors: the metric, the extrinsic curvature, and its
covariant derivatives.  In order to find the equilibrium (\ie, energy
minimizing) shape, one is interested in how $H$ responds to a
deformation of the surface described by a change in the embedding
functions, $\VECX \rightarrow \VECX + \delta \VECX$.  The
straightforward (but tedious) way to do this is to track the course of
the deformation on $\VECX$ through $g_{ab}$, $\sqrt{g}$, $K_{ab}$, and
any appearing covariant derivatives using the structural relationships
(\ref{eq:structuralrelationsship1}) and
(\ref{eq:structuralrelationsship2}).

Alternatively, one can treat $g_{ab}$, $K_{ab}$, $\VECe_{a}$ and
$\VECn$ as \emph{independent} variables, enforcing the structural
relations (\ref{eq:structuralrelationsship1}) and
(\ref{eq:structuralrelationsship2}) using Lagrange multiplier
functions \cite{Guven04}.  One thus introduces the new functional
$H_{\text{c}}[g_{ab},K_{ab},\ldots,\VECX,\VECe_{a},
\VECn,\lambda^{ab},\Lambda^{ab},\VECf^{a},\lambda_{\perp}^{a},\lambda_{n}]$
given by
\begin{eqnarray}
  H_{\text{c}} & = & H[g_{ab},K_{ab}, \ldots] \nonumber \\
  & & \stackrel{\text{(\ref{eq:gab})}}{+} \int \romd A \;
      \lambda^{ab}(g_{ab} - \VECe_{a} \cdot
  \VECe_{b}) \nonumber \\
  & & \stackrel{\text{(\ref{eq:Kab})}}{+}  \int \romd A \;
      \Lambda^{ab}(K_{ab}-\VECe_{a} \cdot \partial_{b} \VECn) \nonumber \\
  & & \stackrel{\text{(\ref{eq:ea})}}{+}  \int \romd A \;
      \VECf^{a} \cdot (\VECe_{a}-\partial_{a}\VECX) \nonumber \\
  & & \stackrel{\text{(\ref{eq:ea_normal_n})}}{+}  \int \romd A \;
      \lambda_{\perp}^{a} (\VECe_{a} \cdot
  \VECn)\nonumber \\
  & & \stackrel{\text{(\ref{eq:n2=1})}}{+}  \int \romd A \;
      \lambda_{n}(\VECn^{2}-1) \ .
  \label{eq:surfacefunctionalnew}
\end{eqnarray}
The original Hamiltonian $H$ is now treated as a function of the
\emph{independent} variables $g_{ab}$, $K_{ab}$ and its covariant
derivatives; $\lambda^{ab}$, $\Lambda^{ab}$, $\VECf^{a}$,
$\lambda_{\perp}^{a}$ and $\lambda_{n}$ are Lagrange multipliers
fixing the constraints (\ref{eq:structuralrelationsship1}) and
(\ref{eq:structuralrelationsship2}). The introduction of auxiliary
variables greatly simplifies the variational problem, because now
we do not have to track explicitly how the deformation $\delta
\VECX$ propagates through to $g_{ab}$ and $K_{ab}$.  As we will see
in the following, this approach also provides a very simple and direct
derivation of the shape equation in which the multiplier $\VECf^a$,
which pins the tangent vectors to the surface, is identified as the
surface stress tensor.

Note that additional physical constraints can be enforced by
introducing further Lagrange multipliers (such as a pressure $P$ in
the case that a fixed volume is enclosed by the surface).

\subsection{Euler-Lagrange equations and the existence of a conserved current \label{subsec:EL}}

The Hamiltonian (\ref{eq:surfacefunctional}) is invariant under
translations.  As explained in detail in
Ref.~\cite{surfacestresstensor}, Noether's theorem then guarantees the
existence of an associated \emph{conserved current}, which we will
identify as the surface stress tensor in
Sec.~\ref{subsec:stresstensor}.  In order to see this, let us first
work out the Euler-Lagrange equations for $\VECX$, $\VECe_{a}$,
$\VECn$, $g_{ab}$ and $K_{ab}$:
\begin{subequations}
  \label{eq:EulerLagrangeequations}
  \begin{eqnarray}
    \nabla_{a} \VECf^{a} & = & 0 \ , \label{eq:ELEinX}
    \\
    \VECf^{a} & = & (\Lambda^{ac} K_{c}^{b} + 2 \lambda^{ab}) \VECe_{b}
      -\lambda_{\perp}^{a}\VECn \ , \label{eq:ELEinea}
    \\
    0 & = & (\nabla_{b}\Lambda^{ab}+\lambda_{\perp}^{a})\VECe_{a}
      +(2\lambda_{n}-\Lambda^{ab}K_{ab})\VECn \ , \label{eq:ELEinn}
    \\
    \lambda^{ab} & = & \textstyle\frac{1}{2}\,T^{ab} \ , \label{eq:ELEingab}
    \\
    \Lambda^{ab} & = & -\mathcal{H}^{ab} \ . \label{eq:ELEinKab}
  \end{eqnarray}
\end{subequations}
Note that the Weingarten equations $\partial_{a}\VECn =
K_{a}^{b}\VECe_{b}$ have been used in Eqn.~(\ref{eq:ELEinea}); the
Gauss equations $\nabla_{a}\VECe_{b}=-K_{ab}\VECn$ have been used in
Eqn.~(\ref{eq:ELEinn}). We have also defined
\begin{subequations}
\begin{eqnarray}
  \mathcal{H}^{ab} & = & \frac{\delta\mathcal{H}}{\delta K_{ab}} \qquad \text{and} \label{eq:Hab} \\
  T^{ab} & = & -\frac{2}{\sqrt{g}}\frac{\delta (\sqrt{g}\mathcal{H})}{\delta g_{ab}} \
  . \label{eq:Tab}
\end{eqnarray}
\label{eq:HabandTab}
\end{subequations}
The manifestly symmetric tensor $T^{ab}$ is the intrinsic stress
tensor associated with the metric $g_{ab}$.  If $\mathcal{H}$ does not
depend on derivatives of $K_{ab}$, functional derivatives in the
definition of $\mathcal{H}^{ab}$ and $T^{ab}$ reduce to ordinary ones.

Equation~(\ref{eq:ELEinX}) reveals the existence of a conservation law
for the current $\VECf^a$.  Using the other equations
(\ref{eq:ELEinn}), (\ref{eq:ELEingab}) and (\ref{eq:ELEinKab}), it is
straightforward to eliminate the Lagrange multipliers on the right
hand side of Eqn.~(\ref{eq:ELEinea}) to obtain an explicit expression
for $\VECf^{a}$ in terms of the original geometrical variables.  From
Eqn.~(\ref{eq:ELEinn}) we find $\lambda_{\perp}^{a} = -\nabla_{b}
\Lambda^{ab}$ because $\VECe_{a}$ and $\VECn$ are linearly
independent; the Eqns.~(\ref{eq:ELEingab}) and (\ref{eq:ELEinKab})
determine $\lambda^{ab}$ and $\Lambda^{ab}$. Thus
Eqn.~(\ref{eq:ELEinea}) can be recast as
\begin{equation}
  \VECf^{a} = (T^{ab}- \mathcal{H}^{ac} K_{c}^{b}) \VECe_{b}
      -(\nabla_{b} \mathcal{H}^{ab}) \VECn \ .
  \label{eq:stresstensorcondeq}
\end{equation}
Once the Hamiltonian density has been specified,
Eqn.~(\ref{eq:stresstensorcondeq}) determines the conserved current
$\VECf^a$ completely in terms of the geometry.  Several representative
examples are treated in the Appendix.

Finally, as pointed out in Refs.~\cite{surfacestresstensor, Guven04},
the \emph{normal projection} of $\nabla_a\VECf^a$ is the
Euler-Lagrange derivative $\mathcal{E}(\mathcal{H})$ of the original
Hamiltonian $H$ which vanishes for an equilibrium shape
\cite{E(H)=P}.  Using the Gauss equations once more, we obtain the
remarkably succinct result
\begin{equation}
\VECn\cdot\nabla_a\VECf^a =
 \mathcal{E}(\mathcal{H}) =
-K_{ab}T^{ab}+(K_{ac}K_b^c-\nabla_a\nabla_b)\mathcal{H}^{ab} \ .
\label{eq:EL_general}
\end{equation}

\subsection{Identification of the stress tensor \label{subsec:stresstensor}}

We will now show that $\VECf^a$ can be identified with the surface stress
tensor.  The variation of the Hamiltonian has a bulk part proportional to
the Euler-Lagrange derivative (\ref{eq:EulerLagrangeequations}) as well as
boundary terms:  under a change in the embedding functions
$\VECX \to \VECX + \delta \VECX$ one gets
\begin{equation}
  \delta H_{\text{c}} = \int \romd A\; \big[\nabla_a \VECf^a \cdot \delta\VECX
    - \nabla_a (\VECf^a \cdot \delta\VECX) \big]
  \label{eq:var}
  \ .
\end{equation}
Additional boundary contributions stem from the variations with
respect to $\VECn$, $g_{ab}$ and $K_{ab}$, since these terms do or may
contain further derivatives which then need to be removed by partial
integration.  However, the one appearing in Eqn.~(\ref{eq:var}) is the
only one that is relevant for identifying the stress tensor: As we
will see below, for this we are exclusively interested in
\emph{translations}, for which $\VECn$, $g_{ab}$ and $K_{ab}$ remain
unchanged.

\begin{figure}
\includegraphics[scale=0.7]{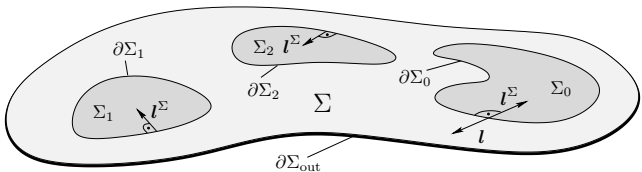}
  \caption{Surface $\Sigma$ with 3 disjoint boundary components
  $\partial\Sigma_i$ and an outer limiting boundary
  $\partial\Sigma_{\text{out}}$.}\label{fig:identification}
\end{figure}

Consider, in particular, a surface region $\Sigma$ \emph{in
equilibrium} (see Fig.~\ref{fig:identification}): its boundary
$\partial\Sigma$ consists of $n$ disjoint closed components
$\partial\Sigma_i$ and an outer limiting boundary
$\partial\Sigma_{\text{out}}$. Each of the $\partial\Sigma_i$ is also
the closed boundary of a surface patch $\Sigma_i$. Under a constant
translation $\delta\VECX=\VECa$ of $\partial\Sigma_i$ the only
non-zero term is
\begin{equation}
  \delta H_{\text{c}} = -\VECa \cdot \oint_{\partial\Sigma_i} \romd s \; l^{\Sigma}_a \VECf^a
    = -\VECa \cdot \VECF^{(i)}_{\rm ext}  \ .
\end{equation}
Stokes' theorem has been used to convert the surface integral into a
line integral.  The vector $\VECl^{\Sigma}=l^{\Sigma}_a\VECe^a$ is the
\emph{outward pointing} unit normal to the boundary on the surface
$\Sigma$; by construction it is tangential to $\Sigma$.  The variable
$s$ measures the arc-length along $\partial\Sigma_i$.  The boundary
integral is thus identified as the external force $\VECF^{(i)}_{\rm
ext}$ acting on $\partial\Sigma_i$: dotted into any infinitesimal
translation, it yields (minus) the corresponding change in energy
\cite{force_at_infinity}.

The external force $\VECF_{\rm ext}$ on the surface patch $\Sigma_0$
is simply given by $-\VECF^{(0)}_{\rm ext}$ due to Newton's third law
\begin{equation}
  \VECF_{\rm ext}
  = \oint_{\partial\Sigma_0} \romd s \; l_a \VECf^a
  = \int_{\Sigma_0} \romd A \; \nabla_a \VECf^a \; ,
  \label{eq:totalforce}
\end{equation}
where $\VECl=-\VECl^{\Sigma}$ and Stokes' theorem was used again.

Recall now that in classical elasticity theory \cite{LaLi_elast} the
divergence of the stress tensor at any point in a strained material
equals the external force density.  Or equivalently, the stress tensor
contracted with the normal vector of a local fictitious area element
yields the \emph{force per unit area} transmitted through this area
element. Comparing this with Eqn.~(\ref{eq:totalforce}) we see that
$\VECf^a$ is indeed the surface analog of the stress tensor:
$l_a\VECf^a$ is the \emph{force per unit length} acting on the
boundary curve due to the action of surface stresses.

It proves instructive to look at the tangential and normal projection
of the stress tensor by defining
\begin{equation}
\VECf^a = f^{ab}\VECe_b + f^a\VECn \ . \label{eq:f_projection}
\end{equation}
Using the equations of Gauss and Weingarten \cite{GaussWeingarten},
the relation $\nabla_a\VECf^a=\mathcal{E}\VECn$ can then be cast in
the form
\begin{subequations}
\label{eq:divf_projections}
\begin{eqnarray}
\nabla_af^a & = & K_{ab}f^{ab} + \mathcal{E} \ ,
\label{eq:divf_normal} \\
\nabla_af^{ab} & = & -K_a^bf^a \ . \label{eq:divf_tangential}
\end{eqnarray}
\end{subequations}
Tangential stress acts as a source of normal stress -- and vice versa.
Both conditions hold irrespective of whether the Euler-Lagrange
derivative $\mathcal{E}$ actually vanishes.  In fact,
Eqn.~(\ref{eq:divf_normal}) shows that the shape equation $\mathcal{E}=0$
is equivalent to $\nabla_af^a=K_{ab}f^{ab}$, while
Eqn.~(\ref{eq:divf_tangential}) merely provides consistency conditions
on the stress components.  For instance, the Helfrich Hamiltonian
$\mathcal{H}\propto K^2$ yields $f^a\propto\nabla^a K$, while $f^{ab}$
is a quadratic in the extrinsic curvature tensor (see
Eqn.~(\ref{eq:stresstensorKnapp})).  Hence,
Eqn.~(\ref{eq:divf_normal}) immediately reproduces the characteristic
form of the Euler-Lagrange derivative: $\Delta K$ plus a cubic in the
extrinsic curvature.

\section{Internal degrees of freedom}\label{sec:tilt}

So far we have restricted the discussion to Hamiltonians which are
exclusively constructed from the geometry of the underlying surface.
However, the surface itself may possess internal degrees of freedom
which can couple to each other and, more interestingly, also to the
geometry.  The simplest example would be a scalar field $\phi$ on the
membrane, which could describe a local variation in surface tension or
lipid composition, and it is readily incorporated into the present
formalism \cite{CaGu04}.

Here we will look a little more closely at the case of an additional
\emph{tangential surface vector field} $m^a$.
Such a field has been introduced to describe the \emph{tilt} degrees
of freedom of the molecules within a lipid bilayer, to accommodate the
fact that the average orientation of the lipids themselves need not
coincide with the local bilayer normal (see for instance
Refs.~\cite{HelPro88,MacLub,NePo9293,SeShNe96,HamKoz00,tilt_force}).
Many additional terms for the energy emerge in the presence of a new
field $m^a$ (for a systematic classification see
Ref.~\cite{NePo9293}). However our aim here is not to treat the most
general case. Instead, we will focus on a simple representative
example to illustrate how easily the present formalism generalizes to
treat such situations.

Let us define the properly symmetrized covariant tilt-strain tensors
$M^{ab}$ and $F^{ab}$ according to
\begin{subequations}
\begin{eqnarray}
M^{ab} & = & \frac{1}{2}\big(\nabla^am^b+\nabla^bm^a\big) \ , \\
\label{eq:Mab}
F^{ab} & = & \;\;\;\;\; \nabla^am^b- \nabla^bm^a \ .
\label{eq:Fab}
\end{eqnarray}
\end{subequations}
In the spirit of a harmonic theory we construct a Hamiltonian density
$\mathcal{H}_\romm$ from the following quadratic invariants:
\begin{equation}
\mathcal{H}_\romm
=
\frac{1}{2}\lambda M^2
+\mu M_{ab}M^{ab}
+\frac{1}{4}\nu F_{ab}F^{ab}
+ V(m^2) \ ,
\label{eq:materialHamiltonian}
\end{equation}
where $M=g_{ab}M^{ab}=\nabla_am^a$ is the tilt divergence.  The first
two terms coincide with the lowest order intrinsic terms identified by
Nelson and Powers \cite{NePo9293}, provided we restrict to unit
vectors $m^a$ \cite{unit_vectors}.  These terms are multiplied by new
elastic constants $\lambda$ and $\mu$, playing the analogous role to
Lam\'{e}-coefficients \cite{Lame}.  If $m^2 \ne 1$ a third term (also
absent in usual elasticity theory \cite{LaLi_elast}) occurs, the
quadratic scalar constructed from the antisymmetrized tilt gradient;
its structure is completely analogous to the Lagrangian in
electromagnetism \cite{LaLi_electro}.  Finally, if the magnitude of
$m^a$ is not fixed, we may also add a potential $V$ depending on the
square $m^2=m_am^a$ of the vector field $m^a$.  Without loss of
generality we assume that $V(0)=0$, because any nonvanishing constant
is more appropriately absorbed into the surface tension $\sigma$.  If
$V(x)$ is minimal for $x=0$, then $m^a\equiv 0$ will minimize the
energy, but depending on physical conditions $V$ may favor nonzero
values of $|m^a|$.  This is why below the main phase transition
temperature of lipid bilayers the lipids can acquire a spontaneous
tilt.

This particular choice for $\mathcal{H}_\romm$ is purely intrinsic.
Hence, Eqn.~(\ref{eq:stresstensorcondeq}) shows that the corresponding
material stress $\VECf_\romm^a$ is also purely intrinsic, therefore
tangential, and given by $\VECf_\romm^a=T^{ab}_\romm\VECe_b$, where
$T^{ab}_\romm=-2\sqrt{g}^{-1}\delta(\sqrt{g}\mathcal{H}_\romm)/\delta
g_{ab}$ is the metric material stress. Performing the functional
variation (see Appendix) we find
\begin{eqnarray}
T^{ab}_\romm & = & \frac{1}{2}\Big[\lambda\big(M^2+2m^c\nabla_cM\big)
+\nu\big(\varepsilon_{cd}\nabla^cm^d\big)^2\Big]g^{ab}
\nonumber \\
& & + \, \mu\Big[-M_{cd}M^{cd}g^{ab} + 2MM^{ab}+2m^c\nabla_cM^{ab}
\nonumber \\
& & \qquad \, -\,(\nabla_cm^a)(\nabla^cm^b) +(\nabla^am_c)(\nabla^bm^c)
\Big]
\nonumber \\
& & -\, V(m^2)g^{ab}-2V'(m^2)m^am^b \ ,
\label{eq:materialmetricstress}
\end{eqnarray}
where $\varepsilon_{ab} = \VECn\cdot(\VECe_a\times\VECe_b)$ is the
antisymmetric epsilon-tensor \cite{epsilon_tensor}.  Notice that the
metric stress tensor is \emph{quadratic} in the tilt-strain, not
linear.  Unlike the stress tensor in elasticity theory, this tensor is not
obtained as the derivative of the energy with respect to the strain
but rather with respect to the \emph{metric}, which leaves it
quadratic in the strain.  The formal analogy alluded to earlier is
therefore not complete.

Adding the material stress $T^{ab}_\romm$ to the tangential
\emph{geometric} stress $f^{ab}$, we find with the help of
Eqns.~(\ref{eq:divf_projections}) the equilibrium conditions
\begin{subequations}
\label{eq:EL_with_tilt}
\begin{eqnarray}
0 & = & -K_{ab}T^{ab}_\romm + \mathcal{E} \ , \\
\nabla_aT^{ab}_\romm & = & 0 \label{eq:EL_with_tilt:material}\ .
\end{eqnarray}
\end{subequations}
The first of these equations shows how the material degrees of freedom
``add'' to the geometric Euler-Lagrange derivative $\mathcal{E}$ of
the geometric Hamiltonian $\mathcal{H}$; this is the modified shape
equation.  The second equation -- which before provided consistency
conditions on the geometrical stresses -- tells us that the material
stress tensor is conserved. The equilibrium of the material degrees of
freedom involves the vanishing of the Euler-Lagrange derivative with
respect to the field $m^a$, which is given by
\begin{eqnarray}
\mathcal{E}_{\romm\,a} \; = \;
\frac{\delta H_\romm}{\delta m^a} & = &
-\lambda \nabla_a\nabla_b m^b
-(\mu+\nu) \nabla_b\nabla_a m^b
\nonumber \\
& & -\; (\mu-\nu) \Delta m_a + 2V'(m^2)m_a \ . \;\;\;
\label{eq:EL_m}
\end{eqnarray}
In general, the equilibrium condition $\mathcal{E}_{\romm\,a}\equiv 0$ implies
Eqn.~(\ref{eq:EL_with_tilt:material}). For a single vector field
$m^a$ the converse also holds so that
Eqn.~(\ref{eq:EL_with_tilt:material}) may be used in place of
the equilibrium condition \cite{equiv}.

In equilibrium, we have not only $\mathcal{E}_{\romm\,a}\equiv 0$,
we also have $\nabla^a\mathcal{E}_{\romm\,a}=0$.  Using the commutation relations
for covariant derivatives \cite{commutation}, it is then easy to see
that the tilt also satisfies the following equation on the surface:
\begin{equation}
(\lambda+2\mu)\Delta M + \mu\nabla^a(Rm_a) - 2\big[2V''m^2+V'M\big] \; = \; 0 \ .
\label{eq:divE}
\end{equation}
Notice that $\nu$ has dropped out of this equation, which follows from
the fact that $F^{ab}$ is invariant under $U(1)$ gauge transformations
\cite{gauge}.  For small values of tilt, we can expand the potential
as
\begin{equation}
V(m^2) = \frac{1}{2}tm^2 + \frac{1}{4}um^4 + \cdots
\label{eq:V_expansion}
\end{equation}
In the untilted phase we can terminate this expression after the first
term (since then $t>0$).  If we now restrict to a flat membrane (and
thus $R\equiv 0$) Eqn.~(\ref{eq:divE}) simplifies to a Helmholtz
equation for the tilt divergence:
\begin{equation}
\big[(\lambda+2\mu)\Delta - t \big]\,M \; = \; 0 \ ,
\label{eq:Helmholtz}
\end{equation}
showing that (in lowest order) any nonzero $M$ is (essentially)
exponentially damped with a decay length of
\begin{equation}
  \ell_\romm = \sqrt{\frac{\lambda+2\mu}{t}} \ .
\label{eq:ell_m}
\end{equation}
If $t<0$ gets us into the tilted phase, the expansion
(\ref{eq:V_expansion}) has to be taken one order higher, leaving
instead a nonlinear Ginzburg-Landau equation to be solved.

We finally remark that even though the system of Euler-Lagrange
equations (\ref{eq:EL_with_tilt}) is quite formidable, it still enjoys
one nice nontrivial property: The material equation
(\ref{eq:EL_with_tilt:material}) is purely intrinsic.  This is the
case because the material stress is tangential, which itself derives
from the fact that the material Hamiltonian is intrinsic.  If we were
to add a coupling between tilt and extrinsic curvature, such as the
chiral term $\varepsilon_{ac}K^c_bm^am^b$, this decoupling would no
longer hold.

\section{Forces between particles \label{sec:forcesbetweenparticles}}

Particles bound to an interface can exert indirect forces onto each
other.  Since these are mediated by the interface, they must be
encoded in its geometry.  We have seen that the ``coding'' is done by
the surface stress tensor $\VECf^a$.  The problem is to decode this
content.

In this section we will solve this problem.  The strong link between
stress and geometry can be easily turned into \emph{exact} expressions
for mediated interactions.  The method by which we obtain these
results for various different Hamiltonians as well as the final
formulas are one of the major results of this paper.

\subsection{The stress tensor and external forces \label{subsec:generalforce}}

Consider a single simply-connected  patch $\Sigma_0$. The external
force acting on it is given by Eqn.~(\ref{eq:totalforce}).
If there are no \emph{external} \cite{external} forces acting on
$\Sigma_{0}$, the integrals appearing in
Eqn.~(\ref{eq:totalforce}) will vanish; but even when $\VECF_{\rm
ext}$ does not vanish, the stress tensor remains divergence free
(Eqn.~(\ref{eq:ELEinX})) on any part of the surface not externally
acted upon. As a result, the contour integral appearing in
Eqn.~(\ref{eq:totalforce}) will be independent of the particular
closed curve so long as it continues to enclose the source of
stress and does not encroach on any other sources.

Observe now that in general a multi-particle configuration can be
stationary only if external forces constrain the particle
positions. These are the forces providing the source of stress in
Eqn.~(\ref{eq:totalforce}). The force $\VECF$ we are ultimately
interested in is the force on a particle mediated by the interface
\emph{counteracting} this external force; we therefore evidently
have $\VECF = - \VECF_{\rm ext}$.

\subsection{Force between particles on a fluid membrane \label{subsec:forceHelfrich}}

Let us now focus on a symmetric fluid membrane, described by the
surface Hamiltonian
\begin{equation}
  \mathcal{H} = \frac{1}{2}\kappa \, K^{2} + \sigma \ ,
  \label{eq:Hamiltoniangeneral}
\end{equation}
which, up to irrelevant boundary terms, is equivalent to the
Hamiltonians introduced by Canham \cite{Canham} and Helfrich
\cite{Helfrich}. Here, $\kappa$ is the bending rigidity and
$\sigma$ is the lateral tension imposed on the boundary. For
typical phospholipid membranes $\kappa$ is of the order of a few
tens of $k_{\text{B}}T$, where $k_{\text{B}}T$ is the thermal
energy. Values for $\sigma$ are found to be in a broad range
between 0 up to about 10 mN/m \cite{surfacetension}. The
Hamiltonian~(\ref{eq:Hamiltoniangeneral}) covers interesting
special cases in various limits: soap films on setting $\kappa=0$
and tensionless membranes on setting $\sigma=0$. Note that the two
elastic constants provide a characteristic length
\begin{equation}
  \ell:=\sqrt{\frac{\kappa}{\sigma}} \ ,
  \label{eq:characteristiclength}
\end{equation}
separating short length scales over which bending energy dominates
from the large ones over which tension does.

We now need to determine the force~(\ref{eq:totalforce}) on a particle
for the Hamiltonian described by Eqn.~(\ref{eq:Hamiltoniangeneral}).
Using Eqns.~(\ref{eq:stresstensor_surfacetension}) and
(\ref{eq:stresstensorKnapp}) from the Appendix, we obtain
\begin{equation}
  \VECf^a
  =
  \Big[\kappa\big(K^{ab}-\frac{1}{2}Kg^{ab}\big)K - \sigma g^{ab}\Big]\,\VECe_b
  - \kappa(\nabla^aK)\,\VECn
  \label{eq:stresstensorHelfrich}
\end{equation}
for the surface stress tensor associated with this Hamiltonian. To
facilitate the calculation of the force it is convenient to introduce
an orthonormal basis of tangent vectors $\{\VECt,\VECl\}$ adapted to
the contour $\partial\Sigma_{0}$: $\VECt = t^a \VECe_a$ points along
the integration contour and, as introduced previously, $\VECl = l^a
\VECe_a $ points normally outward.  The elements of the
extrinsic curvature tensor with respect to this basis are given by
\begin{subequations}
\begin{eqnarray}
  K_\perp & = & l^a l^b K_{ab} \ , \\
  K_{\|} & = & t^a t^b K_{ab} \ , \\
  K_{\perp\|} & = & l^{a}t^{b}K_{ab} \ .
\end{eqnarray}
\end{subequations}
We obtain for the integrand appearing in the line integral in
Eqn.~(\ref {eq:totalforce}),
\begin{equation}
  l_{a} \VECf^{a}
   =
  \Big[\kappa\big(l_{a}K^{ab}-\frac{1}{2}Kl^{b}\big)K-\sigma l^{b}\Big]\,\VECe_b
  - \kappa(\nabla_\perp K)\,\VECn  \ ,
  \label{eq:lf}
\end{equation}
where we have defined the normal derivative $\nabla_\perp = l_a
\nabla^a$. The first term can be simplified by exploiting the
completeness of the tangent basis,
$g_{b}^{c}=l_{b}l^{c}+t_{b}t^{c}$:
\begin{eqnarray}
  l_{a}K^{ab}\VECe_b & = & l_{a}K^{ab} (l_{b}l^{c}+t_{b}t^{c}) \VECe_{c}
  \nonumber \\
  & = & l_{a}l_{b}K^{ab}\VECl +l_{a}t_{b}K^{ab}\VECt \nonumber
  \\
  & = & K_{\perp} \VECl + K_{\perp\|} \VECt \ .
\end{eqnarray}
Since furthermore the trace $K=K_{\perp}+K_{\|}$, we find
\begin{eqnarray}
  \VECF & = & - \oint_{\partial\Sigma_{0}} \!\!\! \romd s \;
    \bigg\{
    \Big[\frac{1}{2}\kappa\big(K_\perp^{2}-K_{\|}^{2}\big)-\sigma\Big]\VECl
     \nonumber
  \\
  && \qquad\qquad
     + \; \kappa K_{\perp\|} K \VECt
     - \kappa\big(\nabla_\perp K\big)\VECn\bigg\}
  \ .
  \label{eq:lf1}
\end{eqnarray}
Note that the integrand has been decomposed with respect to a
(right-handed) orthonormal basis adapted to the contour,
$\{\VECl,\VECt,\VECn\}$.

\subsection{Two-particle configurations \label{subsec:twoparticles}}

\begin{figure}
\includegraphics[scale=0.33]{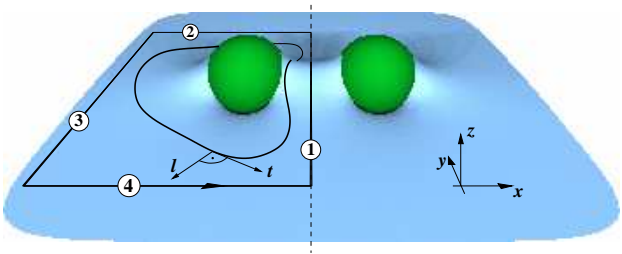}
  \caption{Two identical particles bound to an interface. As described
  in the text, the contour of integration can be deformed in order to
  take advantage of available symmetries.}\label{fig:schematic}
\end{figure}

We are interested in applying the general considerations of
Sec.~\ref{subsec:generalforce} to surface mediated interactions between
colloidal particles.
In particular, we will consider a symmetrical configuration consisting of two identical
particles bound to an asymptotically flat surface, as sketched schematically in
Fig.~\ref{fig:schematic}.

We label by $\{\VECx,\VECy,\VECz\}$ the Cartesian basis vectors of
three-dimensional Euclidian space $\RR^{3}$. Remote from the
particles, the surface is parallel to the $(x,y)$ plane.

Let us agree that the constraining force fixes only the
\emph{separation} between the particles; their \emph{height}, as
well as their \emph{orientation} with respect to the $(x,y)$ plane
are free to adjust and thus to equilibrate. This is also true of
the contact line between surface and colloid when it is not
pinned.  Indeed, Kim \etal\ \cite{Kim} carefully argue that
vertical forces and horizontal torques typically exceed horizontal
forces and vertical torques by a significant amount.  Since the
former can thus be assumed to very quickly equilibrate, they
generally do not contribute to the membrane mediated interaction.

\begin{figure}
\includegraphics[scale=0.9]{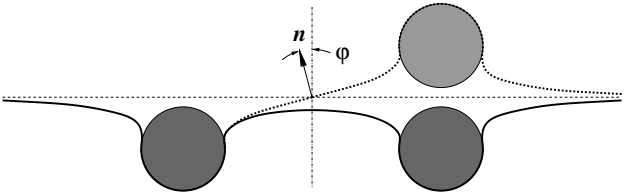}
  \caption{Cross-section of a symmetric (solid line) and an
  antisymmetric (dotted line) two-particle geometry.
  }\label{fig:symmetry}
\end{figure}

There are two distinct manifestations of two-particle symmetry in
this situation:  either a mirror symmetry in the $(y,z)$ plane
(the \emph{symmetric} case) or a twofold symmetry axis, coinciding
with the $y$ axis (the \emph{antisymmetric} case). The former is
relevant if the two particles adhere to the same side of the
surface, the latter applies if they adhere on opposite sides (see
Fig.~\ref{fig:symmetry}). In these two geometries the line joining
corresponding points on the two particles
lies along the $x$-direction.

It is now possible to deform the contour of the line
integral~(\ref{eq:totalforce}) to our advantage: as indicated in
Fig.~\ref{fig:schematic}, the contour describing the force on the left
hand particle may always be pulled open so that the surface is flat on
three of its four branches (2, 3 and 4).  The contributions from
branch 2 will cancel that from 4; the only mathematically involved
term stems from branch 1. The force on the particle is then given by
\begin{equation}
  \VECF = -
  \bigg[\int_{1} + \int_{3}\bigg] \romd s \; l_{a}\VECf^{a} \ .
  \label{eq:totalforce13}
\end{equation}
Let us now apply this general approach to a surface whose energetics can be
described by the Hamiltonian density~(\ref{eq:Hamiltoniangeneral}).

\subsection{The force between particles with symmetry}\label{subsec:explicitforceformulas}

\subsubsection{Fluid membranes}\label{subsubsec:Helfrichcase}

Both mirror and two-fold axial symmetry of branch 1 imply that in
Eqn.~(\ref{eq:lf1}) the tangential term proportional to $\VECt$
vanishes.  In the first case this follows from the fact that branch 1
becomes a line of curvature; hence, the curvature tensor is diagonal
in $(\VECl,\VECt)$-coordinates and thus $K_{\perp\|}$ vanishes.  In
the second case two-fold axial symmetry forces both $K_{\|}$ as well
as $K_\perp$ to be zero, since branch 1 becomes a straight line and
the profile is antisymmetric. In consequence, $K=K_\perp+K_{\|}=0$.
We thereby obtain the first important simplification of the force from
Eqn.~(\ref{eq:lf1}) on that branch:
\begin{equation}
  \VECF_{1}
  =
  -\int_{1} \romd s \; \bigg\{
    \Big[\frac{1}{2}\kappa\big(K_\perp^{2}-K_{\|}^{2}\big)-\sigma\Big]\VECl
    -\kappa\big(\nabla_{\perp} K\big)\VECn\bigg\} \ .
  \label{eq:forcebranch1}
\end{equation}
We now examine separately the two symmetric geometries (see
discussion in Sec.~\ref{subsec:twoparticles}).

\emph{a. Symmetric case.} Tangent and normal vector on branch
1 lie in the $(y,z)$-plane, hence $\VECl=\VECx$. The derivative of $K$
in the direction of $\VECl$ along branch 1, $\nabla_{\perp}K$, is zero
due to mirror symmetry. On branch 3 the surface is flat and thus the
stress tensor is equal to $\VECf_{a,3} = -\sigma\VECe_a$. With this
information we can calculate the total force $\VECF_1 + \VECF_3 =
F_{\text{sym}}\VECx$ on the particle:
\begin{equation}
  F_{\text{sym}}
  \; = \;
  \sigma \Delta L -
    \frac{1}{2} \kappa \int_{1} \romd s \; \big(K_\perp^{2}-K_{\|}^{2}\big) \ ,
  \label{eq:generalforceformula1}
\end{equation}
where $\Delta L\ge 0$ is the excess length of branch 1 compared to
branch 3. If $\kappa=0$, we immediately have the important general
result that the force is always attractive irrespective of the
detailed nature of the source. Unfortunately, the curvature
contribution has no evident sign in general. However, for two
\emph{parallel cylinders} adhering to the \emph{same side} of the
interface the overall sign becomes obvious, as long as the
particles are long enough such that end effects can be neglected:
the contribution $K_{\|}^2$ then vanishes because branch 1 becomes
a line. For the same reason $\Delta L = 0$. This leads to the
formula
\begin{equation}
  F_{\text{sym,cyl}}/L
  \; = \;
  -\frac{1}{2} \kappa K_{\perp}^{2} \ ,
  \label{eq:F.sym.cyl}
\end{equation}
where $L$ is the length of one cylinder. Thus, the two cylinders
repel each other.

\emph{b. Antisymmetric case.} Here branch 1 is a twofold symmetry
axis and, as we have seen above, $K_{\|}=K_\perp=0$. While the
sign of $\nabla_\perp K_{\|}$ is not obvious, the derivative
$\nabla_\perp K_\perp$ is smaller than zero because $K_\perp$
changes sign from positive to negative. The profile on the midline
is always tilted by the angle $\varphi(s)$ in the direction
indicated in Fig.~\ref{fig:symmetry}, because any geometry with
more than one nodal point in the height function between the
particles is expected to possess a higher energy. We fix the
horizontal separation of the particles and allow other degrees of
freedom, such as height or tilt, to equilibrate (see
Sec.~\ref{subsec:twoparticles}). The force on the particle is
therefore parallel to $\VECx$, $\VECF_{\text{antisym}} =
F_{\text{antisym}}\VECx$, and given by
\begin{eqnarray}
  F_{\text{antisym}}
  & = &
  \int_{1} \romd s \; \Big[\sigma\big(\cos{\varphi(s)}-1\big)
  \nonumber \\
  & & \qquad
    - \kappa \sin{\varphi(s)}\,\nabla_{\perp}\big(K_{\perp}+K_{\|}\big)\Big] \ ,
  \label{eq:generalforceformula2}
\end{eqnarray}
where we have used $\VECx \cdot \VECl = \cos{\varphi}$ and
$\VECx \cdot \VECn= -\sin{\varphi}$ at the midpoint. Note that in this case
the tension contribution is repulsive. As before,
the sign of the curvature term is
not obvious.

If we restrict ourselves to the case of two \emph{parallel cylinders}
adhering to \emph{opposite sides} of the interface, however, then
$\nabla_{\perp}K_{\|}$ vanishes at the midpoint.  Furthermore,
$|\VECf_{a}|$ is constant on each of the three free membrane segments
(due to Eqn.~(\ref{eq:ELEinX})). The stress tensor at branch 1,
$\VECf^{\perp}:=\VECf_{a,1}$, must be horizontal to the $\VECx$ axis
because vertical components equilibrate to zero as mentioned above.
Let us look at the projection of the stress tensor onto $\VECl$:
\begin{equation}
  f_{\VECl} := \VECf^{\perp} \cdot \VECl
  = (\VECf^{\perp} \cdot \VECx)(\VECx \cdot \VECl) \ .
\end{equation}
It follows that $\VECf^{\perp}\cdot\VECx =
\text{sign}(f_{\VECl} / \VECx \cdot \VECl) \,
|\VECf^{\perp}|$. We know that $\VECx \cdot \VECl =
\cos\varphi>0$ and $f_{\VECl}=-\sigma<0$. Hence,
$\VECf^{\perp}=-|\VECf^{\perp}|\VECx$ at the midpoint. This reduces
Eqn.~(\ref{eq:generalforceformula2}) to
\begin{equation}
  F_{\text{antisym,cyl}}/L = |\VECf^{\perp}| - \sigma
  = \sqrt{\sigma^2+(\kappa\nabla_\perp K_\perp)^2} - \sigma \ge 0 \ ,
  \label{eq:F.antisym.cyl}
\end{equation}
which implies particle attraction. The length $L$ is again the
length of one cylinder.

\subsubsection{Membranes with tilt degree of freedom \label{subsubsec:degreesoffreedom}}

In Sec.~\ref{sec:tilt} we introduced a tangential vector field
$m^a$ on the membrane, thereby modeling the
degrees of freedom associated with the tilt of the lipids.
The minimal intrinsic Hamiltonian
density Eqn.~(\ref{eq:materialHamiltonian}) already gives
rise to a  quite formidable additional metric stress,
Eqn.~(\ref{eq:materialmetricstress}). Yet, for sufficiently symmetric
situations the expression for the force simplifies quite dramatically,
as we will now illustrate with another striking example.

\begin{figure}
\includegraphics[scale=0.7]{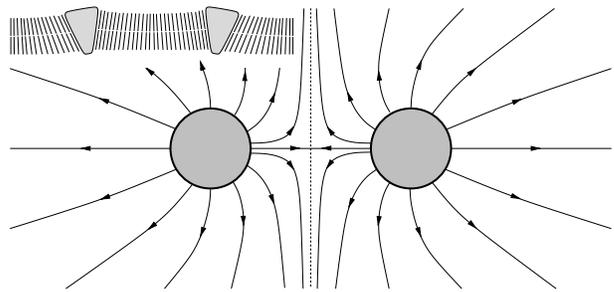}
  \caption{Two conical inclusions act as sources of a local membrane
  tilt (inset).  The tilt-field-lines are illustrated qualitatively
  in this symmetric situation.}\label{fig:splay_illustration}
\end{figure}

Let us consider two conical membrane inclusions which are inserted
with the same orientation into a membrane at some fixed distance apart.
Each inclusion will, due to its up-down-asymmetry, act as a
\emph{local source of tilt}.  Provided the membrane is not in a
spontaneously tilted phase, this tilt will decay with some
characteristic decay length as described at the end of
Sec.~\ref{sec:tilt}.  A typical situation may then look like the one
depicted in Fig.~\ref{fig:splay_illustration}.  What can we say about
the forces between the two inclusions mediated by the tilt field?

Following the same reasoning as for the geometrical forces discussed
above, and remembering that the tilt vanishes on branch 3 so that its
contribution vanishes, we find
\begin{equation}
\VECF_\romm = -\int_1 \romd s \; l_aT_\romm^{ab}\VECe_b \ ,
\end{equation}
with $T_\romm^{ab}$ given by Eqn.~(\ref{eq:materialmetricstress}).  To
simplify this expression, we need to have a close look at the
symmetry.  For this it is very helpful to again expand vectors and
tensors in a local orthonormal frame $(\VECl,\VECt)$, just as we have
done in the geometrical case above.  Mirror symmetry then informs us
that $m^{\|}$ is an \emph{even} function along the direction
perpendicular to branch 1, while $m^\perp$ is an odd function and thus
in particular zero everywhere on that branch.  It thus follows that
both $\nabla_\perp m^{\|}$ and $\nabla_{\|}m^\perp$ vanish everywhere
on branch 1.  Thus we have
\begin{subequations}
\begin{eqnarray}
  M & \stackrel{1}{=} & (\nabla_\perp m^\perp) + (\nabla_{\|}m^{\|}) \ , \\
  M_{ab}M^{ab} & \stackrel{1}{=} &  (\nabla_\perp m^\perp)^2 + (\nabla_{\|}m^{\|})^2 \ , \\
  \varepsilon^{ab}\nabla_am_b & \stackrel{1}{=} & \nabla_\perp m^{\|}-\nabla_{\|}m^\perp \; = \; 0 \ , \label{eq:epsabnabanabb}
\end{eqnarray}
\end{subequations}
where the ``$1$'' above the equation signs reminds us that this only
holds on branch 1.  We next need to look at the contractions of the
individual terms in the metric material stress with $l_a\VECe_b$.  We
find:
\begin{subequations}
\begin{eqnarray}
l_a(\nabla_cm^a)(\nabla^cm^b)\VECe_b & \stackrel{1}{=} & (\nabla_\perp m^\perp)^2\VECl \ , \\
l_a(\nabla^am_c)(\nabla^bm^c)\VECe_b & \stackrel{1}{=} & (\nabla_\perp m^\perp)^2\VECl \ , \\
l_a M M^{ab} \VECe_b & \stackrel{1}{=} & M(\nabla_\perp m^\perp)\,\VECl \ .
\end{eqnarray}
\end{subequations}
The two terms involving the derivatives $m^c\nabla_c$ can be rewritten
by extracting a total derivative:
\begin{subequations}
\begin{eqnarray}
l_a m^c(\nabla_cM)g^{ab}\VECe_b & \stackrel{1}{=} & \VECl \, m^{\|}\nabla_{\|}M
\nonumber \\
& \stackrel{1}{=} &  \VECl \big[ \nabla_{\|}(m^{\|}M) - (\nabla_{\|}m^{\|})M\big] \ . \;\;\;\;\;\;
\end{eqnarray}
The total derivative will yield a boundary term once integrated along
branch 1, and since we assume that we are not in a spontaneously
tilted phase, $|m^a|$ will go to zero at infinity and thus the
boundary term vanishes.  With the same argument we find
\begin{eqnarray}
l_a m^c(\nabla_cM^{ab})\VECe_b & \stackrel{1}{=} &
\VECl \big[\nabla_{\|}(m^{\|}(\nabla_\perp m^\perp)) \nonumber \\
& & - \, (\nabla_{\|}m^{\|})(\nabla_\perp m^\perp)\big] \ .
\end{eqnarray}
\end{subequations}
Again, the total derivative integrates to zero.  Finally, the
potential terms simplify to
\begin{subequations}
\begin{eqnarray}
l_a V(m^2)g^{ab}\VECe_b & \stackrel{1}{=} & V(m^2)\,\VECl \ , \\
l_a V'(m^2)m^am^b\VECe_b & \stackrel{1}{=} & 0 \ .
\end{eqnarray}
\end{subequations}
Collecting all results, we arrive at the remarkably simple exact force
expression $\VECF_\romm = F_\romm\VECx$, with
\begin{eqnarray}
F_\romm & = & -\; \big({\textstyle\frac{1}{2}}\lambda+\mu\big) \int_1\romd s \;
\Big[(\nabla_\perp m^\perp)^2-(\nabla_{\|}m^{\|})^2\Big]
\nonumber \\
& & + \; \int_1\romd s \; V(m^2) \ .
\label{eq:tilt_forces}
\end{eqnarray}
There are two contributions to the force, one stemming from
\emph{gradients} of the tilt, the other from the \emph{tilt potential}
$V$.  Remarkably, the tilt gradient contribution from each of the
first two quadratic invariants has the same structural form, thus the
Lam\'{e} coefficients $\lambda$ and $\mu$ occur only as a combination
in front of the integral.  The modulus $\nu$ has dropped out since the
corresponding stress vanishes on the mid-curve (see
Eqn.~(\ref{eq:epsabnabanabb})).

The structural similarity of Eqn.~(\ref{eq:tilt_forces}) to curvature
mediated forces -- Eqn.~(\ref{eq:generalforceformula1}) -- is very
striking.  Since $\frac{1}{2}\lambda+\mu > 0$ \cite{Lame}, the first
integral states that perpendicular gradients of the perpendicular tilt
lead to repulsion, while parallel gradients of the parallel tilt imply
attractions -- the same ``$\perp^2-\|^2\,$'' motif as found in
Eqn.~(\ref{eq:generalforceformula1}).  Since in the untilted phase
$V(m^2)\ge 0$, the second line shows that the integrated excess
potential drives attraction, just as the excess length (something like
an integrated ``surface tilt'') drives attraction in
Eqn.~(\ref{eq:generalforceformula1}).  Unfortunately, the overall sign
of the force is not obvious.  Looking at the field lines in
Fig.~\ref{fig:splay_illustration}, the visual analogy with
electrostatic interactions between like charged particles would
suggest a repulsion, but the above analysis advises caution (in
Sec.~\ref{subsec:lipid_tilt} we will see that this naive guess is at
least borne out on the linearized level).  Moreover, we should not
forget that tilt \emph{does} couple to geometry (namely, via the
covariant derivative) and that the membrane by no means needs to be
flat; hence, the contribution due to tension and bending given by
Eqn.~(\ref{eq:generalforceformula1}) must be added, the sign of which
is equally unclear.

\subsubsection{Further geometric Hamiltonians \label{subsubsec:otherHamiltonians}}

Within the framework of reparametrization invariant Hamiltonians
providing a scalar energy density, a systematic power series in terms
of all available scalars and their covariant derivatives (each
multiplying some phenomenological ``modulus'') is a formal (and in
fact standard) way of obtaining an energy expression of a physical
system.  In this respect the Hamiltonian (\ref{eq:Hamiltoniangeneral})
is no exception, being simply the quadratic expansion for an up-down
symmetric surface (notice that a term proportional to $K$ would break
this symmetry, giving rise to a spontaneous curvature).  We hasten to
add that a second quadratic term, proportional to the Gaussian (or
Ricci) curvature, exists as well, but this usually plays no role since
it only results in a topological invariant (see also the Appendix).

The fact that curvature (a ``generalized strain'') enters
quadratically in the Hamiltonian (\ref{eq:Hamiltoniangeneral})
classifies this form of the bending energy as ``linear curvature
elasticity'' (even though the resulting shape equations are highly
nonlinear).  However, for sufficiently strong bending higher than
quadratic terms will generally contribute to the energy density,
giving rise to genuinely nonlinear curvature elasticity
\cite{GoeHel96}.  Nevertheless, such effects pose no serious problem
for the approach we have outlined so far. In fact, they are
incorporated very naturally.  We would like to illustrate this with
two examples.

\emph{a. Quartic curvature.} Sticking with up-down symmetric
surfaces, the next curvature order would be quartic, and this
gives rise to three more scalars: $K^4$, $K^2R$ and $R^2$. Let us
for simplicity only study the case of a quartic contribution of the
form
\begin{equation}
  \mathcal{H}_4 = \frac{1}{4}\kappa_{4} K^{4}\ .
  \label{eq:HamiltonianK4}
\end{equation}
Using the general expression of the stress tensor for the scalar
$K^n$ as calculated in the Appendix (see
Eqn.~(\ref{eq:stresstensorKnapp})) and going through the
calculation from Sec.~\ref{subsec:explicitforceformulas} we find
for instance
\begin{equation}
  F_{\text{sym,cyl}}/L = -\frac{3}{4} \kappa_{4} K_{\perp}^{4} \ ,
  \label{eq:F.sym.cyl.K4}
\end{equation}
if two parallel cylinders adhere to the same side of the
interface.  This term increases the repulsion between cylinders
found on the linear elastic level (see Eqn.~(\ref{eq:F.sym.cyl})),
\emph{provided} $\kappa_4>0$, \ie, provided the quartic term
further stiffens the membrane.

Assuming that $\mathcal{H}_4$ perturbs the usual bending
Hamiltonian $\frac{1}{2}\kappa K^2$, we can use the two moduli to
define a characteristic length scale $\ell_4 :=
\sqrt{|\kappa_{4}|/\kappa}$. The overall force up to quartic order
can then be written as
\begin{equation}
  F_{\text{sym,cyl}}/L  =
  -\frac{1}{2}\kappa K_{\perp}^{2} \,
    \Big[1 \pm \frac{3}{2}(\ell_4 K_{\perp})^{2}\Big] \ ,
\end{equation}
where the $+$-sign corresponds to stiffening.  Notice that the
correction term becomes only noticeable once the curvature radius
of the membrane is no longer large compared to the length scale
$\ell_4$. It appears natural that $\ell_4$ is related to the
membrane \emph{thickness}, which for phospholipid bilayers is
about $5\,\text{nm}$. Assuming a (quadratic order) bending
stiffness of $\kappa \simeq 20\,k_{\text{B}}T$, we thus expect
values for the modulus $\kappa_4$ on the order of
$10^3\,k_{\text{B}}T\,\text{nm}^2$.

\emph{b. Curvature gradients.} In order {\sc length}$^{-4}$ it is
possible to also generate scalars which depend on
\emph{derivatives} of the surface curvature.  One such term is
\begin{equation}
H_\nabla = \frac{1}{2}\kappa_\nabla(\nabla_{a}K)(\nabla^{a}K) \ .
\end{equation}
Using the expression for the stress tensor derived in the Appendix
(see Eqn.~(\ref{eq:stresstensor_nabK2})) and again going through
the calculation in Sec.~\ref{subsec:explicitforceformulas}, we
find
\begin{equation}
  F_{\text{sym,cyl}}/L =
    \frac{1}{4} \kappa_\nabla \nabla_{\perp}^{2}K_{\perp}^{2} =
    \frac{1}{4} \kappa_\nabla \frac{\romd^{2}}{\romd l^{2}}K_{\perp}^{2}
\end{equation}
for the force between two symmetrically adhering cylinders. It
depends on very subtle details of the membrane shape:  the
curvature is (roughly) a second derivative of the membrane
position, and this we need to square and differentiate two more
times. Unfortunately, the sign of the interaction is not obvious
here, as the second derivative of $K_{\perp}^{2}$ with respect to
$l$ may be either positive or negative.  Finally, we can also
define a characteristic length scale here, $\ell_\nabla :=
\sqrt{|\kappa_\nabla|/\kappa}$.  The importance of a perturbation
$\mathcal{H}_\nabla$ of the usual bending Hamiltonian depends on
whether or not the curvature \emph{changes} significantly on
length scales comparable to $\ell_\nabla$.


\section{Description of the surface in Monge parametrization \label{sec:Monge}}

In the previous section analytical expressions for the force
between two attached particles  have been derived which link the
\emph{force} to the \emph{geometry} of the surface at the midplane
between them.  It is worthwhile reemphasizing that they are
\emph{exact}, even in the nonlinear regime.  In special cases, the
\emph{sign} of the interaction is also revealed.

If one is interested in quantitative results, however, shape equations
need to be solved -- numerically or analytically.  Either way, one
needs to pick a surface parametrization.  The choice followed in
essentially all existing calculations in the literature is ``Monge
gauge'', and for analytical tractability its linearized version.  The
purpose of this and the following section is to translate the general
covariant formalism developed so far into this more familiar language.
To this end we first remind the reader what the basic geometric
objects look like in this gauge.  We are then in a position to
quantitatively study three different examples of interface mediated
interactions in Sec.~\ref{sec:examples}.

\subsection{Definition and properties \label{subsec:Mongegeneral}}

Any surface free of ``overhangs'' can be described in terms of its
height $h(x,y)$ above some reference plane, which we take to be the
$(x,y)$ plane.  Notice that $x$ and $y$ thus become the surface
coordinates.  The direction of the basis vectors
$\{\VECx,\VECy,\VECz\}\in\RR^{3}$ is as described in
Sec.~\ref{subsec:twoparticles}.

The tangent vectors on the surface are then given by $\VECe_{x} =
(1,0,h_{x})^{\top}$ and $\VECe_{y}=(0,1,h_{y})^{\top}$, where $h_{i} =
\partial_{i}h$ ($i,j\in\{x,y\}$). The metric is given by
\begin{equation}
  g_{ij}= \delta_{ij}+h_{i}h_{j} \ ,
  \label{eq:metricMonge}
\end{equation}
where $\delta_{ij}$ is the Kronecker symbol.  Observe that
$g_{ij}$ is \emph{not} diagonal; even though the coordinates
$\{x,y\}$ refer to an orthonormal coordinate system \emph{on the
base plane}, this property does not transfer to the surface they
parameterize. We also define the gradient operator in the base
plane, $\VECnab := (\partial_{x},\partial_{y})^{\top}$. The metric
determinant can then be written as $g = |g_{ij}| = 1+(\VECnab
h)^{2}$, and the inverse metric is given by $g^{ij} =
\delta_{ij}-h_{i}h_{j}/g$. It is, perhaps, worth emphasizing that
the latter, just as Eqn.~(\ref{eq:metricMonge}), are not tensor
identities. The right-hand side gives the numerical values of the
components of the covariant tensors $g_{ij}$ and $g^{ij}$ with
respect to the coordinates $x$ and $y$.

The unit normal vector is equal to
\begin{equation}
  \VECn=\frac{1}{\sqrt{g}}
    \left(\begin{array}{c} -\VECnab h \\ 1 \end{array}\right) \ .
  \label{eq:normalvectorMonge}
\end{equation}
With the help of Eqn.~(\ref{eq:structuralrelationsship1}) the extrinsic
curvature tensor is determined to be:
\begin{equation}
  K_{ij}= -\frac{h_{ij}}{\sqrt{g}} \ ,
  \label{eq:extrinsiccurvatureMonge}
\end{equation}
where $h_{ij}=\partial_{i}\partial_{j}h$. Note that
Eqn.~(\ref{eq:extrinsiccurvatureMonge}) again is not a tensor equation; it
provides the numerical values of the components of $K_{ij}$ in Monge gauge.

Finally, it is also possible to write the trace $K$ of the extrinsic curvature
tensor in Monge parametrization:
\begin{equation}
  K=-\VECnab \cdot {\Big(\frac{\VECnab h}{\sqrt{g}}}\Big) \ .
  \label{eq:traceofextrinsiccurvatureMonge}
\end{equation}

\subsection{Small gradient expansion \label{subsec:smallgradients}}

In Sec.~\ref{sec:examples} we will be interested in surfaces that
deviate only weakly from a flat plane. In this situation the gradient
$\VECnab h$ is small, and it is sufficient to consider only the lowest
nontrivial order of a small gradient expansion.  $K$ and $\romd A$ can
then be written as
\begin{eqnarray}
  K & = & -\VECnab^{2} h + \mathcal{O}[(\VECnab h)^{2}] \ ,
  \label{eq:traceofextrinsiccurvatureMongesmallgradient} \\
  \romd A & = & {\Big\{1+\frac{1}{2}(\VECnab h)^{2}+
    \mathcal{O}[(\VECnab h)^{4}]}\Big\} \; \romd x \; \romd y \ .
  \label{eq:surfaceelementMongesmallgradient}
\end{eqnarray}

To evaluate the line integrals described in
Sec.~\ref{subsec:explicitforceformulas} we need expressions for
$K_{\perp}$ and $K_{\|}$ as well as the derivatives
$\nabla_{\perp}K_{\perp}$ and $\nabla_{\perp}K_{\|}$ at branch 1
in Monge parametrization.
In the small gradient expansion, the result is simply
\begin{subequations}
  \label{eq:principalcurvaturesMongesmallgradient}
  \begin{eqnarray}
    K_{\perp} & = & - h_{xx}(0,y) \ , \label{eq:KperpMongesmallgradient}
  \\
    K_{\|} & = & - h_{yy}(0,y) \ , \label{eq:KparMongesmallgradient}
  \end{eqnarray}
\end{subequations}
as well as
\begin{subequations}
  \label{eq:principalcurvaturesderperpMongesmallgradient}
  \begin{eqnarray}
    \nabla_{\perp}K_{\perp} & = & - h_{xxx}(0,y) \ ,
    \label{eq:KperpderperpMongesmallgradient}
  \\
    \nabla_{\perp}K_{\|} & = & - h_{yyx}(0,y) \ .
    \label{eq:KparderperpMongesmallgradient}
  \end{eqnarray}
\end{subequations}
We are now in a position to determine the forces between two particles in
different situations.


\section{Examples \label{sec:examples}}

In this section we will illustrate the general framework of
geometry-encoded forces by treating three important examples in Monge
gauge: capillary, curvature mediated, and tilt-induced interactions.
For the first two, force-distance curves have previously been derived
on the linearized level \cite{pinningexp,pinningtheo,Weiklcyl}.  The
route via the stress tensor reproduces these results with remarkable
ease, thereby underscoring its efficiency and also confirming its
validity (at least on the linear level).  To illustrate tilt-mediated
interactions we restrict to a simplified situation in which we neglect
the coupling of membrane shape and tilt-order.  Even if the
\emph{geometry} is ``trivial'' (a flat membrane), the \emph{material}
stress tensor is not, and forces remain.

Both geometric examples are special cases of the Hamiltonian
density~(\ref{eq:Hamiltoniangeneral}).  When the gradients are small,
the surface energy is given by the quadratic expression:
\begin{equation}
  H = \frac{1}{2} \int \romd x \; \romd y \; {\Big[\kappa(\VECnab^{2}h)^{2}
  + \sigma(\VECnab h)^{2} }\Big] \ .
  \label{eq:HamiltoniangeneralMonge}
\end{equation}
If $\kappa=0$ this describes a soap film; if $\kappa \ne 0$ it will
describe a fluid membrane.

The approach, traditionally followed in the literature, is to
first determine the surface profile $h(x,y)$ which minimizes the
energy Eqn.~(\ref{eq:HamiltoniangeneralMonge}).  For this one must
solve the \emph{linear} Euler-Lagrange equation
\begin{equation}
\VECnab^2\big(\VECnab^2-\ell^{-2}\big)\,h(x,y) = 0 \ ,
\label{eq:EL_linear}
\end{equation}
where $\ell$ is the length from
Eqn.~(\ref{eq:characteristiclength}). In a next step, the energy
corresponding to this shape is evaluated by reinserting the solution
of Eqn.~(\ref{eq:EL_linear}) into the functional
(\ref{eq:HamiltoniangeneralMonge}). This energy will depend on the
relative positions of the bound objects.  Appropriate derivatives of
the energy with respect to these positions will yield the forces
between the particles.  By contrast our approach --- sidestepping the
need to evaluate the energy --- will be to determine the force
directly from the surface profile using the line integral expressions
for the force, Eqns.~(\ref{eq:generalforceformula1}) and
(\ref{eq:generalforceformula2}).

\subsection{Soap films \label{subsec:soapfilms}}

For a soap film, $\kappa=0$ and thus $\ell=0$.  The relevant
Euler-Lagrange equation is therefore the Laplace equation,
$\VECnab^2h=0$.

Consider first the \emph{symmetrical} configuration consisting of
two parallel \emph{cylindrical particles} which adhere to one side
of the soap film. Eqn.~(\ref{eq:generalforceformula1}) indicates
that, if we neglect end effects, the force between the cylinders
is proportional to the excess length on branch 1. The excess
length, however, is zero because the contact lines are straight.
Therefore, the force is also zero. Likewise, in the
\emph{antisymmetric} configuration with adhesion on opposite sides,
the soap film between the cylinders will be flat if the vertical
particle displacements are allowed to equilibrate. Therefore,
$\varphi(s)=0$ (see Fig.~\ref{fig:symmetry}) and
Eqn.~(\ref{eq:generalforceformula2}) will yield a zero force exactly
as in the symmetric case. In an analogous way one obtains the same
result for the case of two spheres.

The situation is less simple if the film is \emph{pinned} to the
particle surface. Let us consider \emph{two spherical particles} of
radius $a$ with a contact line that departs only weakly from a circle.

Stamou \etal\ \cite{pinningexp} have studied this case by using a
superposition ansatz in the spirit of Nicolson \cite{nicolson}: first,
the height function of one isolated particle is determined with the
correct boundary conditions. Then, the complete height function is
assumed to be the sum of the two single-particle fields of each of the
two colloids. Strictly speaking, this approach destroys the boundary
conditions at the particles' contact lines; it does, however, give the
correct leading order result for large separation
\cite{exactsolution}.

Using polar coordinates $\rho$ and $\phi$, the solution of the
shape equation outside a single spherical particle can be written
as \cite{pinningexp}
\begin{eqnarray}
  h_{\text{sphere}}(\rho,\phi)
    & = & A_{0} \ln{{\Big(\frac{a}{\rho}}\Big)} \nonumber
  \\
  && + \sum_{m=1}^{\infty} A_{m} \cos{[m(\phi-\phi_{m,0})]}
    {\Big(\frac{a}{\rho}}\Big)^{m} \ , \; \; \; \; \; \; \;
\end{eqnarray}
with multipole coefficients $A_{m}$ and phase angles $\phi_{m,0}$.
The former can be determined as follows: The monopole $A_{0}$
vanishes because there is no external force such as gravity
pulling on the particle. The dipole coefficient $A_{1}$
characterizes the \emph{tilt} of the contact line relative to the
$\VECz$ axis; it also vanishes if there is no external torque
acting on the sphere. All higher multipole coefficients can be
read off from the Fourier expansion of the contact line at
$\rho=a$. It is intuitively obvious and indeed confirmed by a more
careful calculation \cite{pinningexp,pinningtheo} that the
quadrupole dominates the energy at lowest order.

\begin{figure}
\includegraphics[scale=0.50]{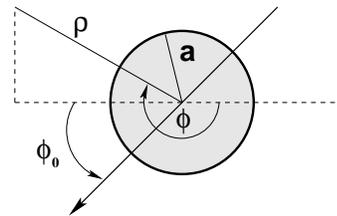}
  \caption{Definition of the coordinates for a single quadrupole
  (viewed from above).}\label{fig:onequadrupole}
\end{figure}

One can therefore restrict the calculation to the single-particle
height function \cite{pinningexp}
\begin{equation}
  h_{\text{sphere}}(\rho,\phi)=Q \cos[2(\phi-\phi_{0})]
    {\Big(\frac{a}{\rho}}\Big)^{2} \ ,
\end{equation}
where $\phi_{0}:=\phi_{2,0}$ is the angle that represents the rotation of the
particle about $\VECz$ (see Fig.~\ref{fig:onequadrupole}).

\begin{figure}
\includegraphics[scale=0.4]{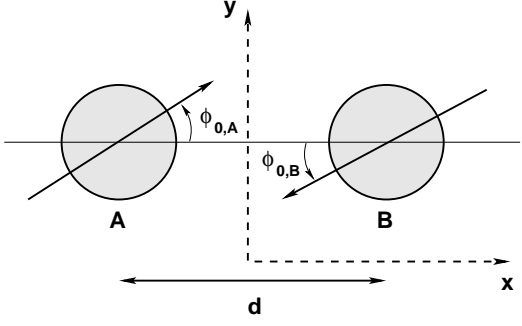}
  \caption{Two quadrupoles on a soap film (viewed from
  above).}\label{fig:twoquadrupoles}
\end{figure}

If the complete height function is a superposition, as described
above, the force on the left particle in lowest order has been
found to be \cite{pinningexp,pinningtheo}
\begin{equation}
  \VECF_{\text{sym,soap}}=-\VECF_{\text{antisym,soap}}=48\pi\sigma Q^{2}
    \frac{a^{4}}{d^{5}}\VECx \ ,
  \label{eq:pinningforce}
\end{equation}
for the symmetric ($\phi_{0,\romA}=-\phi_{0,\romB}$) and the
antisymmetric ($\phi_{0,\romA}=0, \; \phi_{0,\romB}=\pi /2$)
configurations (see Fig.~\ref{fig:twoquadrupoles}).

Let us now examine the same two configurations using the line integral
representation for the force.

\emph{a. Symmetric case.} The force
Eqn.~(\ref{eq:generalforceformula1}) is proportional to the excess
length of branch 1 with respect to branch 3. To quadratic order in
gradients, this difference can be written as
\begin{eqnarray}
  \Delta L & = & \lim_{L\rightarrow\infty}{\Big\{\int_{-L/2}^{L/2} \romd y \;
    {\Big[\sqrt{1+h_{y}^{2}(0,y)}-1}}\Big] \Big\} \nonumber
  \\
  & = & \lim_{L\rightarrow\infty}{\Big\{\int_{-L/2}^{L/2} \romd y \;
    \Big[\frac{1}{2} h_{y}^{2}(0,y)
     + \mathcal{O}[(\VECnab h)^{4}]\Big]}\Big\} \ . \; \; \; \; \; \; \;
  \label{eq:excesslengthsoapfilmpre}
\end{eqnarray}
The height function along the symmetry line between the particles can be
expressed in Cartesian coordinates as:
\begin{equation}
  h(0,y)=2 Q \cos\Big[2\Big(\arctan{\frac{2y}{d}}+\phi_{0,\romA}\Big)\Big]
    \frac{a^{2}}{y^{2}+\frac{d^{2}}{4}} \ .
\end{equation}
Substituting into the second line of
Eqn.~(\ref{eq:excesslengthsoapfilmpre}) gives
\begin{eqnarray}
  \Delta L & = & \lim_{L\rightarrow\infty}
    \Big[96 Q^{2}\frac{a^{4}}{d^{5}}\arctan{\frac{L}{d}}
     + \mathcal{O}(L^{-1})\Big] \nonumber
  \\
  & = & 48\pi Q^{2} \frac{a^{4}}{d^{5}} \ ,
  \label{excesslengthsoapfilm}
\end{eqnarray}
implying via $F_{\text{sym,soap}}=\sigma\Delta L$ the same
force as obtained from energy minimization,
Eqn.~(\ref{eq:pinningforce}). However, it would be fair to say
that we have gained additional information concerning the nature
of this force, missing before. The force is directly proportional
to the length added to the mid-curve as it is stretched. A
\emph{geometrical interpretation} has been provided for the force.
Recall also that this is a non-perturbative result: it does not
depend on the small gradient approximation.

\emph{b. Antisymmetric case.} In this case the horizontal force is
given by Eqn.~(\ref{eq:generalforceformula2}) with $\kappa=0$:
\begin{eqnarray}
  && F_{\text{antisym,soap}} = \sigma \lim_{L\rightarrow\infty}
    {\Big\{\int_{-L/2}^{L/2} \romd y \;
    {\Big[\VECn \cdot \VECz -1}\Big]}\Big\} \nonumber
  \\
  && = \sigma \lim_{L\rightarrow\infty}
    {\Big\{\int_{-L/2}^{L/2} \romd y
    {\Big[\frac{1}{\sqrt{1+h_{x}^{2}(0,y)}} -1}\Big]}\Big\} \nonumber
  \\
  && = \sigma \lim_{L\rightarrow\infty}
    {\Big\{\int_{-L/2}^{L/2} \romd y {\Big[-\frac{1}{2} h_{x}^{2}(0,y)
    + \mathcal{O}[(\VECnab h)^{4}]}\Big]}\Big\} \ . \; \; \; \; \; \; \,
  \label{eq:pinningforceantisympre}
\end{eqnarray}
The height function between the particles is given by
\begin{eqnarray}
  h(x,y) & = & Q a^2 \times \nonumber \\
  & & \hspace*{-5em}
  \bigg\{
    \frac{\cos{[2(\arctan{\frac{y}{\frac{d}{2}+x}})]}}
    {y^{2}+(\frac{d}{2}+x)^{2}}
    -\frac{\cos{[2(\arctan{\frac{y}{\frac{d}{2}-x}})]}}
      {y^{2}+(\frac{d}{2}-x)^{2}}\bigg\} \ ,
\end{eqnarray}
so that
\begin{equation}
  h_{x}(0,y) = -\frac{32 Q a^{2}d(d^{2}-12y^{2})}{(d^{2}+4y^{2})^{3}} \ .
\end{equation}
Inserting this into Eqn.~(\ref{eq:pinningforceantisympre}) yields a
force which again agrees with the one obtained in
Eqn.~(\ref{eq:pinningforce}).

As an example, let us look at colloids with a radius of
$1\,\mu\text{m}$ trapped at the air-water interface
($\sigma\simeq70\,\text{mN}/\text{m}$), which have a pinning
quadrupole of 1\%\ of their radius ($Q\simeq 10\,\text{nm}$).  At a
separation of $3\,\mu\text{m}$ they feel an (attractive or repulsive)
force of $1\,\text{pN}$, and at a separation of about
$16\,\mu\text{m}$ their interaction energy is comparable to the
thermal energy.  These forces are not particularly strong, but they
act over an exceptionally long range.

\subsection{Fluid Membranes \label{subsec:fluidmembranes}}

To describe a fluid membrane, it is necessary to include the bending
energy in Eqn.~(\ref{eq:HamiltoniangeneralMonge}). Let us focus on the
problem of two parallel adhering cylinders which are sufficiently long
so that end effects can be neglected (the fluid membrane analogue of
the problem examined for soap films). In this case the height function
of the surface depends only on one variable, $x$.  Recall that for the
corresponding soap film case no interaction occurred (in the absence
of pinning), see Sec.~\ref{subsec:soapfilms}.

\emph{a. Symmetric case.} Using the energy route, Weikl
\cite{Weiklcyl} shows that, at lowest order in the small gradient
expansion, the energy per unit length of the cylinder is
\cite{go_to_zero}
\begin{equation}
  E_{\text{sym,cyl}}(d)
  = -\frac{(\kappa +2R^{2}U)^{2}(\tanh \frac{d}{2 \ell}-1)}
    {4\sqrt{\sigma \kappa}R^{2}} \ .
  \label{eq:energycylindersym}
\end{equation}
Here $R$ is the cylinder radius, $U$ is the adhesion energy per
unit area, $\ell$ is the characteristic length defined in
Eqn.~(\ref{eq:characteristiclength}), and $d$ is the distance
between the two centers of the cylinders. To obtain the force per
unit length $L$ of the left cylinder, we differentiate
Eqn.~(\ref{eq:energycylindersym}) with respect to $d$
\cite{sign_of_force}:
\begin{equation}
  F_{\text{sym,cyl}}/L
  = -\frac{1}{2}\kappa\bigg(\frac{\kappa +2R^{2}U}{2\kappa R\cosh\frac{d}{2 \ell}}\bigg)^2 \ .
  \label{eq:forcecylindersym}
\end{equation}
The cylinders always repel.

We would now like to determine the force using the line integral of the
corresponding stress tensor. Rewriting the relevant Eqn.~(\ref{eq:F.sym.cyl}) in
small gradient expansion yields (see Eqn.~(\ref{eq:KperpMongesmallgradient})):
\begin{equation}
  F_{\text{sym,cyl}}/L  \; = \;
  -\frac{1}{2} \kappa h_{xx}^{2}(0) \ .
  \label{eq:F.sym.cyl.smallgradient}
\end{equation}
We use the expression for $h$ given in Ref.~\cite{Weiklcyl}:
\begin{equation}
  h(x) = \frac{(\kappa+2R^{2}U)\cosh{\frac{x}{\ell}}}
    {2\sigma R \cosh{\frac{d}{2 \ell}}} + \text{const} \ .
\end{equation}
Its second derivative with respect to $x$ at $x=0$ is
\begin{equation}
  h_{xx}(0) = \frac{\kappa+2R^{2}U}{2\kappa R \cosh{\frac{d}{2 \ell}}} \ .
\end{equation}
Inserting this result into Eqn.~(\ref{eq:F.sym.cyl.smallgradient}) reproduces
the force given by Eqn.~(\ref{eq:forcecylindersym}).

\emph{b. Antisymmetric case.} For two cylinders on opposite sides
of the membrane the energy is given by \cite{Weiklcyl,go_to_zero}
\begin{equation}
  E_{\text{antisym,cyl}}(d)
  = -\frac{(\kappa +2R^{2}U)^{2}(\coth \frac{d}{2 \ell}-1)}
    {4\sqrt{\sigma \kappa}R^{2}} \ ,
  \label{eq:energycylinderantisym}
\end{equation}
which gives a force on the left cylinder \cite{sign_of_force}
\begin{equation}
  F_{\text{antisym,cyl}}/L
  = \frac{1}{2}\kappa\bigg(\frac{\kappa +2R^{2}U}{2\kappa R\sinh\frac{d}{2 \ell}}\bigg)^2 \ .
  \label{eq:forcecylinderantisym}
\end{equation}
The small gradient expansion of Eqn.~(\ref{eq:F.antisym.cyl}) is
\begin{equation}
  F_{\text{antisym,cyl}}/L =
    \frac{1}{2}\kappa\,(\ell\nabla_{\perp} K_\perp)^{2}
  \stackrel{(\ref{eq:KperpderperpMongesmallgradient})}{=}
    \frac{1}{2}\kappa\,[\ell h_{xxx}(0)]^{2} \ .
  \label{eq:F.antisym.cyl.smallgradient}
\end{equation}
If we now again take the height function from Ref.~\cite{Weiklcyl} we arrive at
\begin{equation}
  h(x) = \frac{(\kappa +2R^{2}U) \sinh{\frac{x}{\ell}}}
    {2\sigma R\sinh{\frac{d}{2 \ell}}} \ ,
\end{equation}
which yields
\begin{equation}
  \ell h_{xxx}(0) = \frac{\kappa +2R^{2}U}
    {2\kappa R\sinh{\frac{d}{2 \ell}}} \ .
\end{equation}
Inserting this into Eqn.~(\ref{eq:F.antisym.cyl.smallgradient})
reproduces the result~(\ref{eq:forcecylinderantisym}).

How big are these forces?  As an example, let us look at an actin
filament ($R\simeq 4\,\text{nm}$) adsorbed onto a membrane with a
typical bending stiffness $\kappa\simeq 20\,k_\romB T$, where $k_\romB
T$ is the thermal energy.  Noting that $\sqrt{2U/\kappa}$ will be the
contact curvature at the point where the membrane detaches from the
adsorbing filament \cite{SeLi90} and that this should not be too much
smaller than the bilayer thickness in order for a Helfrich treatment
to be permissible, we take $2UR^2/\kappa\simeq 1$ as an upper limit.
We then find that two adsorbed actin filaments at a distance
$d\simeq\ell$ (where approximately $\sinh\approx\cosh\approx 1$) feel
a force of about $2-3\,\text{pN}/\text{nm}$.  Alternatively, we can
calculate at what distance the interaction energy per persistence
length of the filament ($\ell_\romp\approx 15\,\mu\text{m}$) is of
order $k_\romB T$.  Using a typical value for cell membranes of
$\ell\approx 50\,\text{nm}$, we obtain a separation of about
$0.7\,\mu\text{m}$.  This is huge, and should remind us of the fact
that on this scale a lot of membrane fluctuations will occur which we
have neglected.  Still, it shows rather vividly that membrane mediated
forces can be very significant.

\subsection{Lipid tilt \label{subsec:lipid_tilt}}

The discussion in Sec.~\ref{sec:tilt} shows that lipid tilt order,
described by the surface vector field $m^a$, influences the shape of
the membrane, even if the Hamiltonian density does not contain an
explicit coupling of $m^a$ to the extrinsic curvature.  The coupled
system of differential equations (\ref{eq:EL_with_tilt}) poses a
formidable task, clearly exceeding the already substantial one for the
undecorated shape equation alone.

Our priority is to illustrate the workings of the general formalism,
therefore we will limit the discussion to a simple case where the
analytical treatment is rather transparent: we will assume that the
membrane itself remains flat, such that the energy density stems
exclusively from lipid tilt (as described by $\mathcal{H}_\romm$ from
Eqn.~(\ref{eq:materialHamiltonian})).  This is not a self-consistent
approximation, but should give a good description in the limit in
which the tilt moduli $\lambda$ and $\mu$ are significantly ``softer''
than the bending modulus.  In this case the inclusions we have talked
about in Sec.~\ref{subsubsec:degreesoffreedom} will predominantly
excite tilt and not bend.  More sophisticated (analytical and
numerical) studies of lipid tilt and mediated interactions exist,
which provide better quantitative answers \cite{tilt_force}.

For flat membranes, the Euler-Lagrange equation (\ref{eq:EL_m})
reduces to
\begin{equation}
(\lambda+\mu)\VECnab\VECnab\cdot\VECm + \mu \VECnab^2 \VECm - 2V'\VECm = 0 \ ,
\label{eq:EL_m_flat}
\end{equation}
where $\VECm$ is the 2d tilt vector in the membrane plane.  Focusing
first on one inclusion, the situation acquires cylindrical symmetry.
Writing $\VECm(\VECr)=m(r)\VECe_r$ and restricting to the untilted
membrane phase, for which the tilt potential is sufficiently well
represented by $V(m^2)=\frac{1}{2}tm^2$ with $t>0$,
Eqn.~(\ref{eq:EL_m_flat}) reduces to a simple Bessel equation
\begin{equation}
  x^2m'' + xm' - (x^2+1)m = 0 \ ,
\end{equation}
where $x=r/\ell_\romm$, $\ell_\romm$ is the length defined in
Eqn.~(\ref{eq:ell_m}), and the prime denotes a derivative with respect
to $x$.  The solution is
\begin{equation}
  m(r) = m_0 \frac{K_1(r/\ell_\romm)}{K_1(r_0/\ell_\romm)} \ ,
  \label{eq:tilt_one_particle_solution}
\end{equation}
where $r_0$ is the radius of the inclusion, $m_0$ the value of the
tilt at this point, and $K_\nu$ a modified Bessel function of the
second kind \cite{abramowitz}.  As anticipated, the tilt decays
essentially exponentially with a decay length of $\ell_\romm$.

Obtaining the exact tilt field for two inclusions is very difficult,
since satisfying the boundary conditions is troublesome.  However, if
we again use the Nicolson approximation \cite{nicolson} and assume
that the total tilt distribution is given by the superposition of two
solutions of the kind (\ref{eq:tilt_one_particle_solution}), things
become manageable.  The tilt-mediated force between two symmetric
inclusions is then obtained by inserting the appropriate values and
derivatives of the tilt field $\VECm(x,y)$ on the mid-line into
Eqn.~(\ref{eq:tilt_forces}).  After some straightforward calculations
we get the force \cite{integral}
\begin{eqnarray}
F_\romm & = & 4t\ell_\romm m_0^{\ast 2} \int_{d^\ast}^\infty\romd \xi \; \frac{1}{\xi\sqrt{\xi^2-d^{\ast 2}}} \times
\nonumber \\
& &
\Big[(\xi^2-2d^{\ast 2})K_0(\xi)K_2(\xi) + (\xi^2-d^{\ast 2})K_1^2(\xi)\Big]
\nonumber \\
& =  &
-2\pi t\ell_\romm m_0^{\ast 2} \, K_1(d/\ell_\romm) \ .
\label{eq:tilt_force}
\end{eqnarray}
where $m_0^\ast=m_0/K_1(r_0/\ell_\romm)$, $d^\ast=d/2\ell_\romm$, and
$d$ is the separation between the inclusions.  As we see, the force is
\emph{repulsive} and decays essentially exponentially with distance
over a decay length of $\ell_\romm$.  Integrating it, we get the
repulsive interaction potential
\begin{equation}
U_\romm(d) = 2\pi t\ell_\romm^2 m_0^{\ast 2} \, K_0(d/\ell_\romm) \ .
\end{equation}

Let us try to make a very rough estimate of how big such a force might
be.  For this we need to obtain some plausible values for the numbers
entering into Eqn.~(\ref{eq:tilt_force}).  For $t$ we may use the
equipartition theorem and argue that $\frac{1}{2}t\langle m^2\rangle a
= \frac{1}{2}k_\romB T$, where $a$ is the area per lipid and $k_\romB
T$ the thermal energy.  Assuming that the root-mean-square
fluctuations of $m$ are $10^\circ$ and using the typical value
$a\simeq 0.75\,\text{nm}^2$, we get $t\simeq 40\,k_\romB
T/\text{nm}^2$.  Assuming further a rather conservative tilt decay
length of the order of the bilayer thickness, \ie\ $\ell_\romm\simeq
5\,\text{nm}$, that the inclusion has a radius of $r_0\simeq
3\,\text{nm}$ and imposes there a local tilt of $m\simeq 0.2$, we find
that two inclusions at a distance of $10\,\text{nm}$ feel a
significant force of about $17\,\text{pN}$.  And at a distance of
$d\approx 22\,\text{nm}$ their mutual potential energy is $1\,k_\romB
T$ compared to the separated state.  Notice that this is much larger
than the Debye length in physiological solution, which is typically
only $1\,\text{nm}$.  Hence, tilt-mediated forces can compete with
more conventional forces, such as (screened) electrostatic
interactions. It should be kept in mind, however, that if we permit the membrane to
bend, some of the tilt strain can be relaxed, thereby lowering the
energy.


\section{Conclusions \label{sec:conclusions}}

We have shown how the stress tensor can be used to relate the forces
between particles bound to an interface directly to the interface
geometry. In this approach, the force on a particle is given by a line
integral of the stress along any closed contour surrounding the
particle. The stress depends only on the local geometry; thus the
force is completely encoded in the surface geometry in the
neighborhood of the curve.

The relationship between the force and the geometry provided by the
line integral is exact. In the linear regime, as we have shown for
selected examples in the previous section, the force determined by
evaluating this line integral reproduces the result obtained by the
more familiar energy based approach. Unlike the latter, however, our
approach permits us to consider large deformations.  The expression
for the line-integral is fully covariant, involving geometrical
tensors; one is not limited to any one particular parametrization of
the surface such as the Monge gauge.  Indeed, as we have seen the
geometrical origins of the force can get lost in this gauge.

As we have emphasized previously, this approach is not a substitute
for solving the nonlinear field equations. To extract numbers, we
\emph{do} need to solve these equations. But even before this is done,
the line integral expression can provide valuable qualitative
information concerning the nature of the interactions between
particles. This is because the geometry along the contour is often
insensitive to the precise conditions binding the particle to the
interface. This contrasts sharply with the energy based approach;
there, one needs to know the entire distribution of energy on the
interface before one can say anything about the nature of the
interaction. As we have seen in the context of a symmetrical
two-particle configuration, it is sometimes relatively easy to
identify qualitative properties of the \emph{geometry}; it is
virtually impossible to make corresponding statements about the
\emph{energy} outside the linear regime.

The stress tensor approach also has the virtue of combining seamlessly
with any approach we choose, be it analytical or numerical, to determine
the surface shape. Thus, for instance, one can find surfaces that
minimize a prescribed surface energy functional using the program
``Surface Evolver'' \cite{Brakke}. The evaluation of the force via
a \emph{line} integral involving the geometry along the contour is
straightforward; in contrast, the evaluation of the energy involves
a \emph{surface} integral, and the forces then follow by a subsequent
\emph{numerical differentiation}. In other words, the route via the
energy requires one more integration but also one more
differentiation.  This appears neither economical nor numerically
robust.

We have illustrated how internal degrees of freedom on the membrane
can be incorporated within this approach using a vector order
parameter describing lipid tilt as an example. It is indeed remarkable
just how readily non-geometrical degrees of freedom can be
accommodated within this geometrical framework. Here again, new exact
non-linear expressions for the force between particles mediated by the
tilt are obtained which are beyond the scope of the traditional
approach to the problem. Various patterns emerge which could not have
been guessed from inspection of the Hamiltonian, in particular the
existence of a ``$\perp^2-\|^2\,$" motif common to the geometrical and
tilt mediated forces between symmetrical particles.

\begin{figure}
  \includegraphics[scale=0.33]{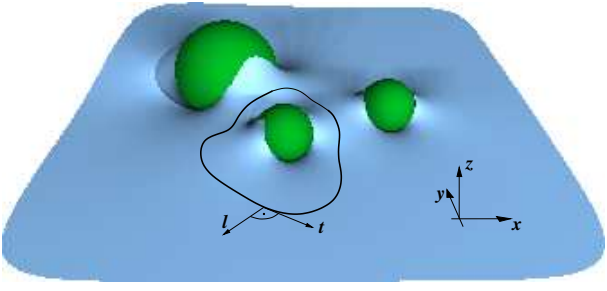}
  \caption{Three-body interactions. The force on one particle can be
  obtained by integrating the surface stress tensor along a line of
  integration enclosing that particle. (cf.\
  Eqn.~(\ref{eq:totalforce})).}\label{fig:multibody}
\end{figure}

We have considered the force between a pair of particles. However, the
interaction between more than two particles is generally not
expressible as a sum over pairwise interactions; superposition does
not hold if the theory is nonlinear (see Ref.~\cite{Kim} for a
striking illustration). This, however, poses no difficulty for the
stress tensor approach, because the underlying relation between
surface geometry and force is independent of whether or not a
pair-decomposition is possible (see Fig.~\ref{fig:multibody}). For
certain symmetric situations a clever choice of the contour of
integration may again yield expressions for the force analogous to
those obtained in Sec.~\ref{subsec:explicitforceformulas}.

Multi-body effects become particularly relevant when one considers
2D bulk phases as, for instance, in a system consisting of a large
number of mutually repulsive particles adhering to one side of an
interface. In this case it is possible to identify expressions
for state variables such as the lateral pressure by exploiting
the approach which has been introduced here.

The interfaces we have considered are asymptotically flat and thus
support no pressure difference. At first glance it may appear that
our approach fails if there is pressure because the stress
tensor is no longer divergence free on the free surface (one has
$\nabla_a\VECf^a=P\VECn$) which would obstruct the deformation of the
contour described in Sec.~\ref{subsec:twoparticles}. As we will show
in a forthcoming publication, however, it is possible to adapt our
approach to accommodate such a situation.

The interactions we have examined correspond to particles whose
orientations are in equilibrium. The additional application of an
external torque (\eg\ on two dipoles via a magnetic field) will
introduce a bending moment. It is, however, possible to treat such
a situation with the tools provided in Ref.~\cite{surfacestresstensor}:
Just as translational invariance gives us the stress tensor,
\emph{rotational} invariance gives us a \emph{torque tensor}.  Its
contour integrals provide us with the torque acting on the patch one
encircles.

Finally, genuine capillary forces involve gravity. The associated
energy, however, depends not only on the geometry of the surface
but also on that of the bulk (it is a volume force). The results
thus differ qualitatively from those presented here.  This will be
the subject of future work.


\begin{acknowledgments}

We would like to thank Riccardo Capovilla for interesting
discussions.  MD acknowledges financial support by the German Science
Foundation through grant De775/1-3. JG acknowledges partial support
from CONACyT grant 44974-F and DGAPA-PAPIIT grant IN114302.
\end{acknowledgments}


\appendix


\section{\label{app:examples}}

In Sec.~\ref{subsec:stresstensor} we derived the general
expression
\begin{equation}
  \VECf^{a} = (T^{ab}- \mathcal{H}^{ac} K_{c}^{\; b}) \VECe_{b}
      -\nabla_{b} \mathcal{H}^{ab} \VECn  \label{eq:stresstensorcondeqapp}
\end{equation}
for the surface stress tensor. In this appendix we will specialize
(\ref{eq:stresstensorcondeqapp}) to a few important standard
cases.

\emph{a. Area.} The simplest case is the area, $\mathcal{H}=1$,
which is (up to a constant prefactor) the Hamiltonian density of a
soap film. We evaluate $\mathcal{H}^{ab}$ and $T^{ab}$ appearing in
Eqn.~(\ref{eq:stresstensorcondeqapp}) using Eqn.~(\ref{eq:HabandTab}):
$\mathcal{H}^{ab} = \delta\mathcal{H}/\delta K_{ab} = 0$ and $T^{ab} =
-g^{ab}$; for $T^{ab}$ we use the identity
\begin{equation}
  \frac{\partial\sqrt{g}}{\partial g_{ab}}=\frac{1}{2}\sqrt{g}\,g^{ab} \ .
  \label{eq:der1nablaKquadrat}
\end{equation}
Thus we get
\begin{equation}
  \VECf^a = -g^{ab}\VECe_b \ .
  \label{eq:stresstensor_surfacetension}
\end{equation}
Note that the functional derivatives $\delta$ in this case are equal
to partial derivatives $\partial$ because $\mathcal{H}$ does not
depend on higher derivatives of $g_{ab}$ or $K_{ab}$.

\emph{b. Powers of $K$.} For the Hamiltonian density $\mathcal{H}
= K^{n} = (g^{ab}K_{ab})^{n}$ one derives \cite{diffapp1}:
$\mathcal{H}^{ab}=n K^{n-1} g^{ab}$ and $T^{ab}=2n K^{n-1} K^{ab}
- K^{n} g^{ab}$ which gives
\begin{equation}
  \VECf^{a}=(nK^{n-1}K^{ab}-K^{n}g^{ab})\VECe_{b}-n (\nabla^{a}K^{n-1}) \VECn
  \ .
  \label{eq:stresstensorKnapp}
\end{equation}
The case $n=2$ is needed in Eqn.~(\ref{eq:Hamiltoniangeneral}).

\emph{c. Einstein-Hilbert action.} Canham \cite{Canham} originally
used the quadratic Hamiltonian $\mathcal{H}=K_{ab}K^{ab}$. For
this one we easily see that $\mathcal{H}^{ab} = 2K^{ab}$ and
$T^{ab} = -\mathcal{H} g^{ab}+4K^a_cK^{bc}$ \cite{diffapp1}. Using
the contractions of both Gauss-Codazzi-Mainardi equations
(\ref{eq:GCM}) as well as the fact that the Euler-Lagrange
derivative $\mathcal{E}(\mathcal{H})$ is linear in the
Hamiltonian, we get with the help of Eqn.~(\ref{eq:EL_general})
the Euler-Lagrange derivative of the Einstein-Hilbert action,
$\mathcal{H}=R$:
\begin{equation}
\mathcal{E}(R) = \mathcal{E}(K^2) - \mathcal{E}(K_{ab}K^{ab}) =
-2K^{ab}G_{ab} \ .
\end{equation}
Here, $G_{ab} = R_{ab}-\frac{1}{2}Rg_{ab}$ is the Einstein tensor,
which vanishes identically in two dimensions.  Thus, surface
variations of $K^2$ and $K_{ab}K^{ab}$ differ only by boundary
terms (in accord with the Gauss-Bonnet theorem
\cite{DifferentialGeometry}). In higher dimensions, however,
$\mathcal{E}(R)\propto G_{ab}$ is a nontrivial result, and the
above seemingly inelegant (since extrinsic) derivation is after
all remarkably economical.

\emph{d. Curvature gradient.} The next example we consider is the
Hamiltonian density $\mathcal{H} = \frac{1}{2}(\nabla_{c}
K)(\nabla^{c} K) \equiv\frac{1}{2}(\nabla K)^2$. Now we need to
keep in mind that $\mathcal{H}^{ab}$ and $T^{ab}$ are functional
derivatives
\begin{equation}
  \mathcal{H}^{ab}=\frac{\delta\mathcal{H}}{\delta K_{ab}}
  =\frac{\partial\mathcal{H}}{\partial K_{ab}}
  -\nabla_{c}{\Big(\frac{\partial\mathcal{H}}{\partial \nabla _{c} K_{ab}}}\Big)
  \ ,
  \label{eq:funcderivativeapp}
\end{equation}
because $\mathcal{H}$ depends on derivatives of $K_{ab}$. We
obtain
\begin{equation}
  \mathcal{H}^{ab}=-\nabla_{c}(g^{ab}\nabla^{c}K)=-g^{ab} \Delta K \ .
  \label{eq:HabnablaKquadratapp}
\end{equation}
The determination of $T^{ab}$ is a little more difficult; to avoid
errors, let us proceed cautiously and consider the variation of
the Hamiltonian $H=\frac{1}{2} \int \romd A \; (\nabla K)^2$ with
respect to the metric $g_{ab}$ and identify $T^{ab}$ at the end of
the calculation. The variation yields:
\begin{eqnarray}
  \delta_{g}H & = &
    \frac{1}{2}\int \romd^{2}\xi \; \delta_{g}[\sqrt{g}(\nabla
    K)^{2}]
  \\
  & = & \frac{1}{2} \int \romd A \;
    \Big\{\frac{\delta_{g}\sqrt{g}}{\sqrt{g}} (\nabla K)^{2}
      + \delta_{g}[(\nabla K)^{2}]\Big\} \ .
  \nonumber
  \label{eq:varHCnablaKquadrat}
\end{eqnarray}
The evaluation of the first term involves the reuse of
Eqn.~(\ref{eq:der1nablaKquadrat}); we expand the second term
\cite{diffapp2}
\begin{eqnarray}
  \delta_{g}[(\nabla K)^{2}]
  & = &
  \delta_{g}[g^{ab}] (\nabla_a K)(\nabla_b K) \nonumber \\
  & &
  + 2 g^{ab} (\nabla_{a}K) \delta_{g}[\nabla_{b}(K_{cd}g^{cd})] \nonumber
  \\[0.5em]
  & = & -(\nabla^{a}K) (\nabla^{b}K) \delta g_{ab} \nonumber \\
  & &
  + 2 g^{ab} (\nabla_{a}K) \nabla_{b}(K_{cd} \delta g^{cd}) \nonumber
  \\[0.5em]
  & = & -(\nabla^{a}K) (\nabla^{b}K) \delta g_{ab} \nonumber \\
  & &
  - 2 g^{ab} (\nabla_{a}K) \nabla_{b}(K^{cd} \delta g_{cd}) \ ,
  \label{eq:term2nablaKquadrat}
\end{eqnarray}
where the identity $\delta_{g} g^{ab} = -g^{ac}g^{bd}\delta
g_{cd}$  is exploited twice. We obtain
\begin{eqnarray}
  \delta_{g}H & = &
  \frac{1}{2} \int \romd A \; \Big[\frac{1}{2}g^{ab}(\nabla K)^{2}
    - (\nabla^{a}K)(\nabla^{b}K)\Big] \delta g_{ab} \nonumber
  \\
  && - \int \romd A \; g^{ab} (\nabla_{a} K) \nabla_{b}(K^{cd} \delta g_{cd})
  \ .
\end{eqnarray}
The last term can be rewritten as
\begin{eqnarray}
  && \int \romd A \; g^{ab} (\nabla_{a} K) \nabla_{b}(K^{cd} \delta g_{cd})
    = - \int \romd A \; K^{ab} \Delta K \delta g_{ab} \nonumber
  \\
  && + \int \romd A \;
    \nabla_{b} {\Big[g^{ab} (\nabla_{a} K) K^{cd} \delta g_{cd}}\Big]
  \ ;
\end{eqnarray}
the second term on the right hand side is a total divergence and
can be cast as a boundary term. Therefore, it does not contribute
to $T^{ab}$.  Collecting results, we find
\begin{equation}
  \delta_{g}H \stackrel{(\ref{eq:Tab})}{=} -\frac{1}{2} \int \romd A \;
    T^{ab} \delta g_{ab} + \text{boundary terms} \ ,
\end{equation}
with
\begin{equation}
  T^{ab} = (\nabla^{a}K) \; (\nabla^{b}K)
    - \frac{1}{2}g^{ab}(\nabla K)^{2}- 2 K^{ab}\Delta K
  \ .
\end{equation}
Thus, for $\mathcal{H}=\frac{1}{2}(\nabla K)^{2}$ we get for
$\VECf^{a}$ given by Eqn.~(\ref{eq:stresstensorcondeqapp}) the
remarkably compact expression
\begin{eqnarray}
  \VECf^{a} & = &
    {\Big[(\nabla^{a}K)(\nabla^{b}K)-\frac{1}{2}g^{ab}(\nabla K)^{2}
      -K^{ab}\Delta K }\Big]\VECe_{b} \nonumber
  \\
    && + \; \nabla^{a}\Delta K \; \VECn \ .
  \label{eq:stresstensor_nabK2}
\end{eqnarray}

\emph{e. Vector field.} As a final example let us consider
Hamiltonians of the kind introduced in Sec.~\ref{sec:tilt}, which have
internal vector degrees of freedom.  With the symmetric tilt-strain
tensor $M^{ab}=\frac{1}{2}(\nabla^am^b+\nabla^bm^a)$ we can for
instance look at the quadratic Hamiltonian density $\mathcal{H} =
\frac{1}{2}(\nabla_a m^a)^2=\frac{1}{2}M^2$, where $m^a$ is the
(contravariant) surface vector field and $M=g_{ab}M^{ab}$. This term
is purely intrinsic, hence $\mathcal{H}^{ab}=0$.  The difficult part
is the covariant differentiation, which acts on a \emph{vector field}
and is thus dressed with an additional Christoffel symbol. Since the
latter depends on the metric and its first partial derivative
\cite{covariant_derivative}, it will contribute to the variation:
\begin{eqnarray}
\delta_g H & = & \frac{1}{2}\int \romd^{2}\xi \;
\delta_{g}[\sqrt{g}M^{2}]
  \nonumber \\
  & = & \frac{1}{2} \int \romd A \;
    \Big\{\frac{\delta_{g}\sqrt{g}}{\sqrt{g}} M^{2}
      + \delta_{g}(m^a_{,a} + \Gamma_{ab}^am^b)^{2}\Big\} \ .
      \;\;\;\;\;\;
  \label{eq:varHCnablaama}
\end{eqnarray}
The first term is once more simplified via
Eqn.~(\ref{eq:der1nablaKquadrat}), while the second term calls for the
Palatini identity \cite{Weinberg}
\begin{equation}
\delta_g \Gamma_{ab}^c = \frac{1}{2}g^{cd}\big[\nabla_b\delta
g_{da}+\nabla_a\delta g_{bd}-\nabla_d\delta g_{ab}\big] \ .
\label{eq:Palatini}
\end{equation}
Since the $\delta g_{ab}$ are the components of a tensor (they must
describe a proper variation of the metric tensor), the variation
$\delta_g\Gamma_{ab}^c$ is also a tensor, even though the Christoffel
symbol itself is not. Using Eqn.~(\ref{eq:Palatini}), the second term
in Eqn.~(\ref{eq:varHCnablaama}) can thus be rewritten as
\begin{eqnarray}
\delta_g(m^a_{,a}+\Gamma_{ab}^am^b)^2 & = & 2Mm^b\delta_g\Gamma_{ab}^a \nonumber \\
& \hspace*{-12em}= & \hspace*{-6em}Mm^bg^{ad}(\nabla_b\delta
g_{da}+\nabla_a\delta
g_{bd}-\nabla_d\delta g_{ab}) \nonumber \\
& \hspace*{-12em}= & \hspace*{-6em}Mm^dg^{ab}(\nabla_d\delta
g_{ab}) \ .
\end{eqnarray}
The derivative of $\delta g_{ab}$ is removed by a final partial
integration.  Collecting everything, we thus find (up to
irrelevant boundary terms)
\begin{eqnarray}
\delta_g H & = & \frac{1}{2} \int \romd A \; \Big\{
\frac{1}{2}M^2-\nabla_d\big[m^dM\big]\Big\}g^{ab}\delta
g_{ab} \nonumber \\
& = & \frac{1}{2} \int \romd A \; \Big\{
-\frac{1}{2}M^2-m^d\nabla_dM\Big\}g^{ab}\delta g_{ab} \ . \;\;\;\;\;\;\;
\end{eqnarray}
Thus, the metric stress tensor is
\begin{equation}
T^{ab} = \frac{1}{2}\Big[M^2+2m^c\nabla_cM\Big]g^{ab} \ .
\end{equation}
Notice that it is directly proportional to the metric; its effect
in the stress tensor will thus be to renormalize the surface
tension.

The second quadratic invariant, $\mathcal{H}=M_{ab}M^{ab}$, does not
provide any additional difficulties compared to $\frac{1}{2}M^2$, even
though the calculation is a bit longer.  One finds:
\begin{eqnarray}
T^{ab} & = & -M_{cd}M^{cd}g^{ab} + 2MM^{ab}+2m^c\nabla_cM^{ab}
\nonumber \\
& & -\,(\nabla_cm^a)(\nabla^cm^b) +(\nabla^am_c)(\nabla^bm^c) \ . \;\;\;
\end{eqnarray}

Finally, the third quadratic invariant $\mathcal{H} =
\frac{1}{4}F_{ab}F^{ab}$ (with $F^{ab}=\nabla^am^b-\nabla^bm^a$)
can be treated rather easily by noting that
$F_{ab}=\partial_am_b-\partial_bm_a$ is independent of the connection.
A short calculation then shows that
\begin{equation}
T^{ab} = F^{ac}F^b_{\;\;c} - \frac{1}{4}g^{ab}F_{cd}F^{cd} \ .
\label{eq:Tab_Fab_general}
\end{equation}
This is nothing but the energy-momentum tensor from electrodynamics
\cite{LaLi_electro}.  In two dimensions it can be further simplified,
since any antisymmetric tensor is then proportional to the
epsilon-tensor: $F^{ab} = \frac{1}{2} \varepsilon^{ab}
\varepsilon_{cd}F^{cd}$.  Inserting this into
Eqn.~(\ref{eq:Tab_Fab_general}) and using the identity
$\varepsilon^{ac}\varepsilon^b_{\;\;c}=g^{ab}$ \cite{epsilon_tensor},
we find
\begin{equation}
T^{ab} = \frac{1}{2}g^{ab}\big(\varepsilon_{cd}\nabla^cm^d\big)^2 \ ,
\end{equation}
showing that the stress is isotropic, just as in the case of the
Hamiltonian $\mathcal{H}=\frac{1}{2}M^2$.  It will thus only
renormalize the surface tension and, in particular, not single out any
specific new directions on the membrane.




\begin{thebibliography}{99}

\bibitem{MTW}
C. W. Misner, K. S. Thorne, and J. A. Wheeler,
\textit{Gravitation}, (W. H. Freeman, New York, 1973);

\bibitem{Wald}
R. M. Wald, \textit{General Relativity}, (University of Chicago Press,
Chicago, 1984).

\bibitem{Weinberg}
S. Weinberg, \textit{Gravitation and Cosmology}, (Wiley, New York,
1972).

\bibitem{Belloni00}
L. Belloni, J. Phys.: Condens. Matter \textbf{12}, R549 (2000).

\bibitem{Likos01}
C. N. Likos, Phys. Rep. \textbf{348}, 267 (2001).

\bibitem{Hill}
T. L. Hill, \textit{An Introduction to Statistical Thermodynamics},
(Dover, New York, 1986);

\bibitem{pinningexp}
D. Stamou, C. Duschl, and D. Johannsmann,
Phys. Rev. E \textbf{62}, 5263 (2000).

\bibitem{pinningtheo}
J.-B. Fournier and P. Galatola,
Phys. Rev. E \textbf{65}, 031601 (2002).

\bibitem{Kralchevsky}
P. A. Kralchevsky, V. N. Paunov, N. D. Denkov, K. Nagayama,
J. Chem. Soc. Faraday Trans. \textbf{91}, 3415 (1995);
%
P. A. Kralchevsky and K. Nagayama, Adv. Coll. Interface
Sci. \textbf{85}, 145 (2000).

\bibitem{Koltover}
I. Koltover, J. O. R{\"a}dler, and C. R. Safinya,
Phys. Rev. Lett. \textbf{82}, 1991 (1999).

\bibitem{Weiklcyl}
T. R. Weikl,
Eur. Phys. J. E \textbf{12}, 265 (2003).

\bibitem{gbp}
M. Goulian, R. Bruinsma, and P. Pincus,
Europhys. Lett. \textbf{22}, 145 (1993);
%
Erratum: Europhys. Lett. \textbf{23}, 155 (1993);
%
note also the further correction in:
J.-B. Fournier and P. G. Dommersnes,
Europhys. Lett. \textbf{39}, 681 (1997).

\bibitem{inclusions}
V. I. Marchenko and C. Misbah,
Eur. Phys. J. E \textbf{8}, 477 (2002);
%
D. Bartolo and J.-B. Fournier,
Eur. Phys. J. E \textbf{11}, 141 (2003);
%
T. R. Weikl, M. M. Kozlov, and W. Helfrich,
Phys. Rev. E \textbf{57}, 6988 (1998).

\bibitem{Fourspher}
P. G. Dommersnes, J.-B. Fournier, and P. Galatola,
Europhys. Lett. \textbf{42}, 233 (1998).

\bibitem{Kim}
K. S. Kim, J. Neu, and G. Oster,
Biophys. J. \textbf{75}, 2274 (1998);
%
K. S. Kim, J. C. Neu, and G. F. Oster,
Europhys. Lett. \textbf{48}, 99 (1999).

\bibitem{BisBis}
P. Biscari, F. Bisi, and R. Rosso,
J. Math. Biol. \textbf{45}, 37 (2002);
%
P. Biscari and F. Bisi,
Eur. Phys. J. E \textbf{7}, 381 (2002).

\bibitem{colloidchargeren}
F. Oosawa, \textit{Polyelectrolytes}, (Dekker, New York, 1971);
%
G. S. Manning, J. Chem. Phys. \textbf{51} 924 (1969);
%
S. Alexander, P. M. Chaikin, P. Grant, G. J. Morales, P. Pincus, and
D. Hone, J. Chem. Phys. \textbf{80}, 5776 (1984).

\bibitem{mem_inter}
M. M. M{\"u}ller, M. Deserno, and J. Guven,
Europhys. Lett. \textbf{69}, 482 (2005).

\bibitem{HelPro88}
W. Helfrich and J. Prost, Phys. Rev. A \textbf{38}, 3065 (1988).

\bibitem{MacLub}
F. C. MacKintosh and T. C. Lubensky, Phys. Rev. Lett. \textbf{67},
1169 (1991);
%
T. C. Lubensky and F. C. MacKintosh, Phys. Rev. Lett. \textbf{71},
1565 (1993).

\bibitem{NePo9293}
P. Nelson and T. Powers, Phys. Rev. Lett. \textbf{69}, 3409 (1992);
J. Phys. II (France) \textbf{3}, 1535 (1993).

\bibitem{SeShNe96}
U. Seifert, J. Shillcock, and P. Nelson, Phys. Rev. Lett. \textbf{77},
5237 (1996).

\bibitem{HamKoz00}
M. Hamm and M. M. Kozlov, Eur. Phys. J. E \textbf{3}, 323 (2000).

\bibitem{tilt_force}
S. May and A. Ben-Shaul, Biophys. J. \textbf{76}, 751 (1999);
J.-B. Fournier, Eur. Phys. J. B \textbf{11}, 261 (1999);
S. May, Eur. Biophys. J. \textbf{29}, 17 (2000);
K. Bohinc, V. Kralj-Igli\v{c}, and S. May,
J. Chem. Phys. \textbf{119}, 7435 (2003);
Y. Kozlovsky, J. Zimmerberg, and M. M. Kozlov, Biophys. J. \textbf{87}
999, (2004).

\bibitem{DifferentialGeometry}
M. Do Carmo,
\textit{Differential Geometry of Curves and Surfaces}, (Prentice Hall,
1976);
%
E. Kreyszig,
\textit{Differential Geometry}, (Dover, New York, 1991).

\bibitem{covariant_derivative}
For instance, for a scalar $\phi$ we have $\nabla_a\phi =
\partial_a\phi$, while for a covariant vector field $m^c$ we have
$\nabla_am^c = \partial_am^c+\Gamma_{ab}^cm^b$, where $\Gamma_{ab}^c$
is the Christoffel symbol of the second kind, defined by
$\Gamma_{ab}^c = \frac{1}{2}g^{cd} (\partial_ag_{bd} +
\partial_bg_{da} - \partial_dg_{ab})$ \cite{DifferentialGeometry}.

\bibitem{Guven04}
J. Guven, J. Phys. A: Math. Gen. \textbf{37}, L313 (2004).

\bibitem{surfacestresstensor}
R. Capovilla and J. Guven,
J. Phys. A: Math. Gen. \textbf{35}, 6233 (2002).

\bibitem{E(H)=P}
For unconstraint surfaces $\mathcal{E}(\mathcal{H}) = 0$ is indeed
the shape equation. However, further constraints imply additional
terms.  For instance, \emph{closed} surfaces yield the equation
$\mathcal{E}(\mathcal{H}) = P$, where depending on the situation
$P$ can be seen as the excess internal pressure or a Lagrange
multiplier conjugate to a volume constraint.  In either case, the
simple conservation law for the current $\VECf^a$ breaks down.

\bibitem{force_at_infinity}
Note that in equilibrium the total force acting on the surface
$\Sigma$ is zero, whereas the different
$\VECF^{(i)}_{\text{ext}}$ do not have to vanish necessarily.

\bibitem{LaLi_elast}
L. D. Landau and E. M. Lifshitz,
\textit{Theory of Elasticity}, (Butterworth-Heinemann, Oxford, 1999).

\bibitem{GaussWeingarten}
The equations of Gauss and Weingarten express changes of the
normal and tangent vectors along the surface in terms of these
vectors. They follow from
Eqns.~(\ref{eq:structuralrelationsship1},
\ref{eq:structuralrelationsship2}) and read, respectively,
$\nabla_a\VECn=K_a^b\VECe_b$ and $\nabla_a\VECe_b=-K_{ab}\VECn$.

\bibitem{CaGu04}
R. Capovilla and J. Guven, J. Phys.: Condens. Matter \textbf{16},
S2187 (2004).

\bibitem{unit_vectors} Using the commutation relations for covariant
derivatives \cite{commutation} it is easy to see that
$(\nabla_am^b)(\nabla_bm^a) = (\nabla_am^a)^2 - \frac{1}{2}Rm^2$.  If
$m^2=1$, the Gauss-Bonnet-theorem renders the last term a boundary
contribution, and it is then easy to see that the $\lambda$- and
$\mu$-terms in Eqn.~(\ref{eq:materialHamiltonian}) are sufficient.

\bibitem{Lame}
In order for the Hamiltonian density~(\ref{eq:materialHamiltonian}) to
be positive definite in the strain gradient $\nabla^am^b$, we need
$\mu>0$ and $\lambda+\mu>0$.  The latter differs from the ``usual''
condition $\lambda+\frac{2}{3}\mu>0$ found in Ref.~\cite{LaLi_elast}
because in the surface case we only have two-dimensional tensors.  The
positivity of the \emph{antisymmetric} term requires $\nu<0$, since
the square of an antisymmetric matrix has negative eigenvalues.

\bibitem{LaLi_electro}
L. D. Landau and E. M. Lifshitz,
\textit{The Classical Theory of Fields}, (Butterworth-Heinemann,
Oxford, 2000).

\bibitem{epsilon_tensor}
We may also write $\varepsilon_{ab} = \sqrt{g}\,\epsilon_{ab}$ with
$\epsilon_{12} = -\epsilon_{21} = 1$ and $\epsilon_{11} =
\epsilon_{22}=0$.  This implies $\varepsilon^{ab} =
\sqrt{g}^{-1}\epsilon_{ab}$.  Another useful identity is
$\varepsilon_{ab}\varepsilon_{cd}=g_{ac}g_{bd}-g_{ad}g_{bc}$.

\bibitem{equiv} The converse is not always
true. Had we considered two vector fields instead of one, $m_1^a$ and $m_2^a$ say, the
corresponding Euler-Lagrange equations $\mathcal{E}_{\romm_1\,a}\equiv 0$, and
$\mathcal{E}_{\romm_2\,a}\equiv 0$ would not follow from the conservation law.
The two sets of Euler-Lagrange equations are required to determine the equilibrium.
The conservation law alone is not enough.


\bibitem{commutation}
Given a contravariant vector $V^c$, we have the commutation identity
$[\nabla_a,\nabla_b]V^c = R_{ab\;\;e}^{\;\;\;\;c}V^e$, for
contravariant tensors of rank 2 a second Riemann tensor arises:
$[\nabla_a,\nabla_b]T^{cd} =
R_{ab\;\;e}^{\;\;\;\;c}T^{ed}+R_{ab\;\;e}^{\;\;\;\;d}T^{ce}$. To
obtain Eqn.~(\ref{eq:divE}), we have also used the fact that for
two-dimensional surfaces the Riemann tensor is given by $R_{abcd}=
\frac{1}{2} R \, \varepsilon_{ab}\varepsilon_{cd} = \frac{1}{2} R (g_{ac}
g_{bd} - g_{ad} g_{bc})$.

\bibitem{gauge} The Euler-Lagrange derivative of the $F^{ab}$
contribution in Eqn.~(\ref{eq:materialHamiltonian}) is easily seen to
be $\mathcal{E}_{\nu\,a} = \nabla^bF_{ab}$.  Furthermore, due to its
asymmetry $F^{ab}$ is invariant under gauge transformations
$m^a\rightarrow m^a+\nabla^a\Lambda$ with some arbitrary gauge
function $\Lambda$.  Performing an infinitesimal transformation, we
get
\begin{equation}
\delta_{\Lambda}\; \frac{1}{4}\int\romd A \, F_{ab}F^{ab} =
-\int\romd A \, (\nabla^a\mathcal{E}_{\nu\,a})\Lambda \ .
\nonumber
\end{equation}
Since $\Lambda$ was arbitrary, $\nabla^a\mathcal{E}_{\nu\,a}$ must
vanish identically.  In the corresponding electrodynamic situation
$\mathcal{E}_{\nu\,a}$ is proportional to the 4-current
\cite{LaLi_electro}; there, the same gauge invariance is responsible
for charge conservation.

\bibitem{external}
Any patch $\Sigma_0$ is connected via its boundary $\partial\Sigma_0$
to its surrounding surface, which will also exert forces onto
$\Sigma_0$ transmitted through this boundary.  This, however, is
\emph{not} what we mean by external forces.  Rather, external forces
originate from \emph{outside} the surface; an example would be
external objects pushing or pulling on the patch $\Sigma_0$.

\bibitem{Canham}
P. B. Canham, J. Theoret. Biol. \textbf{26}, 61 (1970).

\bibitem{Helfrich}
W. Helfrich, Z. Naturforsch. \textbf{28c}, 693 (1973).

\bibitem{surfacetension}
Living cells actively monitor the lateral tension of their plasma
membrane by means of lipid reservoirs which in controlled response
to external stimuli extract or inject lipids into the membrane.
For a review see:
%
C. E. Morris and U. Homann,
J. Membrane. Biol. \textbf{179}, 79 (2001).

\bibitem{GoeHel96}
R. Goetz and W. Helfrich, J. Phys. II France \textbf{6}, 215 (1996).

\bibitem{nicolson}
M. M. Nicolson,
Proc. Cambridge Philos. Soc. \textbf{45}, 288 (1949).

\bibitem{exactsolution}
Note that it is also possible to solve the shape equation exactly in small
gradient expansion \cite{pinningtheo}. However, both calculations yield the same
energy in lowest order. Therefore, the easier approach is presented here.

\bibitem{go_to_zero}
Compared to Ref.~\cite{Weiklcyl} we have shifted the potential such
that it goes to zero as $d \rightarrow \infty$.

\bibitem{sign_of_force}
The \emph{direction} of the force is always opposite for the two particles.
Hence, if we want to encode this information in the \emph{sign}, an additional
minus sign is needed for the left particle, since it moves to the
\emph{negative} $x$-direction upon repulsion and to the positive upon
attraction, respectively.

\bibitem{SeLi90}
U. Seifert, R. Lipowsky, Phys. Rev. A \textbf{42}, 4768 (1990).

\bibitem{abramowitz}
M. Abramowitz and I. A. Stegun, \textit{Handbook of mathematical
functions}, (Dover, New York, 1970).

\bibitem{integral}
Obtaining the integral is straightforward, solving it is not.  We
first obtained the large-$d$-asymptotics of the integral and
recognized this as the large-$d$-asymptotics of the right hand side of
Eqn.~(\ref{eq:tilt_force}).  However, Eqn.~(\ref{eq:tilt_force})
indeed holds for \emph{all} distances $d$, a fact we unfortunately
were only able to verify numerically.

\bibitem{Brakke}
K. Brakke,
\emph{The Surface Evolver},
(http://\linebreak[0]www.susqu.edu/\linebreak[0]facstaff/b/\linebreak[0]brakke/\linebreak[0]evolver/\linebreak[0]evolver.html,
2004).

\bibitem{diffapp1}
One has to be careful when differentiating with respect to
$g_{ab}$: the tensor $K_{ab}$ is an independent variable, hence
$\partial K_{ab}/\partial g_{cd}=0$; but $K^{ab} = K_{cd} g^{ac}
g^{bd}$ depends on the metric through its inverse and thus yields
a nontrivial term when differentiated, since $\partial
g^{ab}/\partial g_{cd} = - \frac{1}{2}(g^{ac}g^{bd}+ g^{ad} g^{bc})$.

\bibitem{diffapp2}
One more warning when differentiating with respect to $g_{ab}$
(\emph{cf.} \cite{diffapp1}): generally, $\nabla_a$ depends on
(derivatives of) the metric through the Christoffel symbols it
contains \cite{covariant_derivative}.  Luckily, however, in the
present situation $\nabla_{a}$ only acts on a \emph{scalar} ($K$). It
is then identical to the usual partial derivative $\partial_a$, \ie,
it is not dressed with additional Christoffel symbols.  On the other
hand, $\nabla^{a}=g^{ab}\nabla_b$ \emph{always} depends on the metric
through its inverse.

\end{thebibliography}
\end{document}